\documentclass[12pt]{iopart}
\usepackage[utf8]{inputenc}
\usepackage[T1]{fontenc}
\expandafter\let\csname equation*\endcsname=\relax
\expandafter\let\csname endequation*\endcsname=\relax
\usepackage{amsmath,amssymb,amsfonts,amsthm}

\catcode`,\active

\catcode`\,12

\newcommand*\pFqskip{8mu}
\catcode`,\active
\newcommand*\pFq{\begingroup
        \catcode`\,\active
        \def ,{\mskip\pFqskip\relax}%
        \dopFq
}
\catcode`\,12
\def\dopFq#1#2#3#4#5{%
        {}_{#1}F_{#2}\biggl[\genfrac..{0pt}{}{#3}{#4};#5\biggr]%
        \endgroup
}
\newcommand{\Keywords}[1]{\par\noindent
{\footnotesize{\textbf{Keywords}\/}: #1}}
\newcommand{\nombrespacs}[1]{\par\noindent
{\small{PACS numbers\/}: #1}}
\newcommand{\class}[1]{\par\noindent
{\small{AMS classification scheme numbers\/}: #1}}
\newcommand{\pd}{\partial}
\newcommand{\ket}[1]{|#1\rangle}
\newcommand{\kket}[2]{|#1\rangle_{#2}}
\newcommand{\bra}[1]{\langle #1|}

\newcommand{\Braket}[3]{\bra{#1}#2\ket{#3}}
\newcommand{\BBraket}[5]{{}_{#1}\bra{#2}#3\ket{#4}_{#5}}
\newcommand{\braket}[2]{\langle #1|#2\rangle}

\newcommand{\under}[1]{_{#1}}
\numberwithin{equation}{section}
\begin{document}
\title[Multivariate Krawtchouk polynomials and $SO(d+1)$ representations]{The multivariate Krawtchouk polynomials as matrix elements of the rotation group representations on oscillator states}
\author{Vincent X. Genest}
\ead{genestvi@crm.umontreal.ca}
\address{Centre de recherches math\'ematiques, Universit\'e de Montr\'eal, C.P. 6128, Succursale Centre-ville, Montr\'eal, Qu\'ebec, Canada, H3C 3J7}
\author{Luc Vinet}
\ead{luc.vinet@umontreal.ca}
\address{Centre de recherches math\'ematiques, Universit\'e de Montr\'eal, C.P. 6128, Succursale Centre-ville, Montr\'eal, Qu\'ebec, Canada, H3C 3J7}
\author{Alexei Zhedanov}
\ead{zhedanov@yahoo.com}
\address{Donetsk Institute for Physics and Technology, Donetsk 83114, Ukraine}
\begin{abstract}
An algebraic interpretation of the bivariate Krawtchouk polynomials is provided in the framework of the $3$-dimensional isotropic harmonic oscillator model. These polynomials in two discrete variables are shown to arise as matrix elements of unitary reducible representations of the rotation group in $3$ dimensions. Many of their properties are derived by exploiting the group-theoretic setting. The bivariate Tratnik polynomials of Krawtchouk type are seen to be special cases of the general polynomials that correspond to particular rotations involving only two parameters. It is explained how the approach generalizes naturally to $(d+1)$ dimensions and allows to interpret multivariate Krawtchouk polynomials as matrix elements of $SO(d+1)$ unitary representations. Indications are given on the connection with other algebraic models for these polynomials.\\

\vfill

\nombrespacs{03.65.Fd, 02.20.-a}\\
\class{06B15, 33C50, 22E46}
\end{abstract}

\pagebreak
\section{Introduction}
The main objective of this article is to offer a group-theoretic interpretation of the multivariable generalization of the Krawtchouk polynomials and to show how their theory naturally unfolds from this picture. We shall use as framework the space of states of the quantum harmonic oscillator in $d+1$ dimensions. It will be seen that the Krawtchouk polynomials in $d$ variables arise as matrix elements of the reducible unitary representations of the rotation group $SO(d+1)$ on the energy eigenspaces of the $(d+1)$-dimensional oscillator. For simplicity, we shall focus on the $d=2$ case. The bivariate Krawtchouk polynomials will thus appear as matrix elements of $SO(3)$ representations; we will indicate towards the end of the paper how the results directly generalize to an arbitrary finite number of variables.

The ordinary Krawtchouk polynomials in one discrete variable have been obtained by Krawtchouk \cite{Krawtchouk-1929} in 1929 as polynomials orthogonal with respect to the binomial distribution. They possess many remarkable properties \cite{Koekoek-2010,Suslov-1991} (second-order difference equation, duality, explicit expression in terms of Gauss hypergeometric function, etc.) and enjoy numerous applications. The importance of these polynomials in mathematical physics is due, to a large extent, to the fact that the matrix elements of $SU(2)$ irreducible representations known as the Wigner $D$ functions can be expressed in terms of Krawtchouk polynomials \cite{Zhedanov-1986,Koornwinder-1982}.

The determination of the multivariable Krawtchouk polynomials goes back at least to 1971 when Griffiths obtained \cite{Griffiths-1971} polynomials in several variables that are orthogonal with respect to the multinomial distribution using, in particular, a generating function method. These polynomials, especially the bivariate ones, were subsequently rediscovered by several authors. For instance, the $2$-variable Krawtchouk polynomials appear as matrix elements of $U(3)$ group representations in \cite{Rosen-1987} and the same polynomials occur as $9j$ symbols of the oscillator algebra in \cite{Zhedanov-1997}. An explicit expression in terms of Gel'fand-Aomoto generalized hypergeometric series is given in \cite{Tanaka-2004}. Interest was sparked in recent years with the publication by Hoare and Rahman of a paper \cite{Rahman-2008} in which the $2$-variable Krawtchouk polynomials were presented, anew, from a probabilistic perspective. This led to bivariate Krawtchouk polynomials being sometimes called Rahman polynomials. A number of papers followed \cite{Grunbaum-2007,Grunbaum-2010,Rahman-2011,Mizukawa-2011}; the approach of \cite{Rahman-2008}, related to Markov chains, was extended to the multivariate case in reference \cite{Grunbaum-2010} to which the reader is directed for an account of the developments at that point in time. 

Germane to the present paper are references \cite{Terwi-2012} and \cite{Iliev-2012}. In the first of these papers, Iliev and Terwilliger offer a Lie-algebraic interpretation of the bivariate Krawtchouk polynomials using the algebra $sl_{3}(\mathbb{C})$. In the second paper, this study was extended by Iliev to the multivariate case by connecting the Krawtchouk polynomials in $d$ variables to $sl_{d+1}(\mathbb{C})$. In these two papers, the Krawtchouk polynomials appear as overlap coefficients between basis elements for two modules of $sl_{3}(\mathbb{C})$ or $sl_{d+1}(\mathbb{C})$ in general. The basis elements for the representation spaces are defined as eigenvectors of two Cartan subalgebras related by an anti-automorphism specified by the parameters of the polynomials. The interpretation presented here is in a similar spirit. We shall indicate in Section 5 and in the appendix what are the main observations that are required if one wishes to establish the correspondence. In essence, the key is in the recognition that the anti-automorphism used in \cite{Iliev-2012} and \cite{Terwi-2012} can be taken to be a rotation (times $i$). The analysis is then brought in the realm of the theory of Lie group representations. This entails connecting two parametrizations of the polynomials: the one used in the cited literature and the other that naturally emerges in the interpretation to be presented, in terms of rotation matrix elements. It is noted that the connection with $SO(d+1)$ rotations readily explains the $d(d+1)/2$ parameters of the polynomials.

A major advance in the theory of multivariable orthogonal polynomials was made by Tratnik \cite{Tratnik-1991}, who defined a family of multivariate Racah polynomials, thereby obtaining a generalization to many variables of the discrete polynomials at the top of the Askey scheme and extending the multivariate Hahn polynomials introduced by Karlin and McGregor in \cite{McGregor-1975} in the context of linear growth models with many types. These Racah polynomials in $d$ variables depend on $d+2$ parameters; they can be expressed as products of single variable Racah polynomials with the parameter arguments depending on the variables. Using limits and specializations, Tratnik further identified multivariate analogs to the various discrete families of the Askey tableau, thus recovering the multidimensional Hahn polynomials of Karlin and McGregor and obtaining in particular an ensemble of Krawtchouk polynomials in $d$ variables depending on only $d$ parameters (in contrast to the $d(d+1)/2$ parameters that we were so far discussing). We shall call these the Krawtchouk-Tratnik polynomials so as to distinguish them from the ones introduced by Griffiths. The bispectral properties of the multivariable Racah-Wilson polynomials defined by Tratnik have been determined in \cite{Iliev-2010}. As a matter of fact, the Krawtchouk-Tratnik polynomials are also orthogonal with respect to the multinomial distribution and have been used in multi-dimensional birth and death processes \cite{Milch-1968}. It is a natural question then to ask what relation do the Krawtchouk-Tratnik polynomials have with the other family. As will be seen, the former are special cases of the latter corresponding to particular choices of the rotation matrix. This fact had been obscured it seems, by the usual parametrization which is singular in the Tratnik case.

To sum up, we shall see that the multivariable Krawtchouk polynomials are basically the overlap coefficients between the eigenstates of the isotropic harmonic oscillator states in two different Cartesian coordinate systems related to one another by an arbitrary rotation. This will provide a cogent underpinning for the characterization of these functions: simple derivations of known formulas will be given and new identities will come to the fore. 

In view of their naturalness, their numerous special properties and especially their connection to the rotation groups, it is to be expected that the multivariable Krawtchouk polynomials will intervene in various additional physical contexts. Let us mention for example two situations where this is so. The bivariate Krawtchouk polynomials have already been shown in \cite{Miki-2012} to provide the exact solution of the $1$-excitation dynamics of a two-dimensional spin lattice with non-homogeneous nearest-neighbor couplings. This allowed for an analysis of quantum state transfer in triangular domains of the plane. The multivariate Krawtchouk polynomials also proved central in the construction \cite{Miki-2013} of superintegrable finite models of the harmonic oscillator where they arise in the wavefunctions. Let us stress that in this case we have a variant of the relation with group theory as the polynomials are basis vectors for representation spaces of the symmetry group in this application. Indeed, the energy eigenstates of the finite oscillator in $d$ dimensions are given by wavefunctions where the polynomials in the $d$ discrete coordinates have fixed total degrees. This is to say that the Krawtchouk polynomials in $d$ variables, with given degree, span irreducible modules of $SU(d)$.

The paper is structured as follows. In Section 2, we specify the representations of $SO(3)$ on the energy eigensubspaces of the three-dimensional isotropic harmonic oscillator. In Section 3, we show that the matrix elements of these representations define orthogonal polynomials in two discrete variables that are orthogonal with respect to the trinomial distribution. In Section 4, we use the unitarity of the representations to derive the duality property of the polynomials. A generating function is obtained in Section 5 using boson calculus and is identified with that of the multivariate Krawtchouk polynomials. The recurrence relations and difference equations are obtained in section 6. An integral representation of the bivariate Krawtchouk polynomials is given in terms of Hermite polynomials in Section 7. It is determined in Section 8 that the representation matrix elements for rotations in coordinate planes are given in terms of ordinary Krawtchouk polynomials in one variable. In Section 9, the bivariate Krawtchouk-Tratnik polynomials are shown to be a special case of the general polynomials associated to rotations expressible as the product of two rotations in coordinate planes. In the group-theoretic interpretation of special functions, addition formulas are the translation of the group product. This is the object of Section 10, in which a simple derivation of the formula expressing the bivariate Krawtchouk-Tratnik polynomials as a product of two ordinary Krawtchouk   polynomials in one variable is given and where an expansion of the general bivariate Krawtchouk polynomials $Q_{m,n}(i,k;N)$ in terms of the Krawtchouk-Tratnik polynomials is provided. We indicate in Section 11 how the analysis presented in details for the two variable case extends straightforwardly to an arbitrary number of variables. It is also explained how the parametrization of \cite{Iliev-2012} is related to the one in terms of rotation matrices. A short conclusion follows. Background on multivariate Krawtchouk polynomials will be found in the Appendix as well as explicit formulas, especially for the bivariate case, relating parametrizations of the polynomials.

\section{Representations of $SO(3)$ on the quantum states of the harmonic oscillator in three dimensions}
In this section, standard results on the Weyl algebra, its representations and the three-dimensional harmonic oscillator are reviewed. Furthermore, the reducible representations of the rotation group $SO(3)$ on the oscillator states that shall be considered throughout the paper are defined.
\subsection{The Weyl algebra}
Consider the Weyl algebra generated by $a_{i}$,  $a_{i}^{\dagger}$, $i=1,2,3$, and defined by the commutation relations
\begin{equation}
\label{Weyl}
[a_i, a_k]=0,\qquad  [a_i^{\dagger}, a_k^{\dagger}]=0, \qquad [a_i, a_k^{\dagger}]=\delta_{ik}.
\end{equation}
The algebra \eqref{Weyl} has a standard representation on the states
\begin{equation}
\label{Basis}
\ket{n_1,n_2,n_3}\equiv \ket{n_1}\otimes \ket{n_2}\otimes\ket{n_3},
\end{equation}
where $n_1$, $n_2$ and $n_3$ are non-negative integers. This representation is defined by the following actions on the factors of the direct product states:
\begin{equation}
\label{Actions}
a_{i}\ket{n_i}=\sqrt{n_i}\,\ket{n_i-1},\quad a_{i}^{\dagger}\ket{n_i}=\sqrt{n_i+1}\,\ket{n_i+1}.
\end{equation}
It follows from \eqref{Actions} that one can write
\begin{equation}
\label{Ket-Definition}
\ket{n_1,n_2,n_3}=\frac{(a_{1}^{\dagger})^{n_1}(a_{2}^{\dagger})^{n_2}(a_{3}^{\dagger})^{n_3}}{\sqrt{n_1!n_2!n_3!}}\ket{0,0,0}.
\end{equation}
The algebra \eqref{Weyl} has a realization in the Cartesian coordinates $x_{i}$ given by
\begin{equation}
\label{Realization}
a_i=\frac{1}{\sqrt{2}}(x_i+\pd_{x_i}),\quad a_{i}^{\dagger}=\frac{1}{\sqrt{2}}(x_i-\pd_{x_i}),
\end{equation}
where $\pd_{x_i}$ denotes differentiation with respect to the variable $x_i$.
\subsection{The 3D quantum harmonic oscillator}
Consider now the Hamiltonian of the three-dimensional quantum harmonic oscillator
\begin{align}
\label{3D-Harmonic-Coordinates}
\mathcal{H}=-\frac{1}{2}\nabla^2+\frac{1}{2}(x_1^2+x_2^2+x_3^2)-3/2.
\end{align}
In terms of the realization \eqref{Realization}, the Hamiltonian \eqref{3D-Harmonic-Coordinates} reads
\begin{align}
\label{3D-Harmonic-Operators}
\mathcal{H}=a_{1}^{\dagger}a_{1}+a_{2}^{\dagger}a_{2}+a_{3}^{\dagger}a_{3}.
\end{align}
From the expression \eqref{3D-Harmonic-Operators} and the actions \eqref{Actions}, it is directly seen that the Hamiltonian $\mathcal{H}$ of the three-dimensional harmonic oscillator is diagonal on the states \eqref{Basis} with eigenvalues $N=n_1+n_2+n_3$:
\begin{align}
\mathcal{H}\ket{n_1,n_2,n_3}=N\ket{n_1,n_2,n_3},
\end{align}
The Schr\"odinger equation
\begin{align*}
\mathcal{H}\Psi=E \Psi,
\end{align*}
associated to the Hamiltonian \eqref{3D-Harmonic-Coordinates} separates in particular in the Cartesian coordinates $x_i$ and in these coordinates the wavefunctions have the expression
\begin{align}
\nonumber
\braket{x_1,x_2,x_3}{n_1,n_2,n_3}&=\Psi_{n_1,n_2,n_3}(x_1,x_2,x_3)\\
&=\frac{1}{\sqrt{2^{N}\pi^{3/2}n_1!n_2!n_3!}}e^{-(x_1^2+x_2^2+x_3^2)/2}H_{n_1}(x_1)H_{n_2}(x_2)H_{n_3}(x_3),
\label{Coordinate-Rep}
\end{align}
where $H_{n}(x)$ stands for the Hermite polynomials \cite{Koekoek-2010}. 
\subsection{The representations of $SO(3)\subset SU(3)$ on oscillator states}
It is manifest that the harmonic oscillator Hamiltonian $\mathcal{H}$, given by \eqref{3D-Harmonic-Coordinates} in the coordinate representation, is invariant under rotations. Moreover, it is clear from the expression \eqref{3D-Harmonic-Operators} that $\mathcal{H}$ is invariant under $SU(3)$ transformations. We introduce the set of orthonormal basis vectors
\begin{align}
\label{Basis-Vectors}
\kket{m,n}{N}=\ket{m,n,N-m-n},\qquad m,n=0,\ldots,N,
\end{align}
which span the eigensubspace of energy $N$ which is of dimension $(N+1)(N+2)/2$. For each $N$, the basis vectors \eqref{Basis-Vectors} support an irreducible representation of the group $SU(3)$, which is generated by the constants of motion of the form $a_{i}^{\dagger}a_{j}$. In the following, we shall however focus on the subgroup $SO(3)\subset SU(3)$, which is generated by the three angular momenta
\begin{align}
\label{Rotation-Generators}
J_{i}=-i\sum_{j,k=1}^{3}\epsilon_{ijk}a_{j}^{\dagger}a_{k},
\end{align}
satisfying the commutation relations
\begin{align*}
[J_i,J_{j}]=i\epsilon_{ijk}J_{k},
\end{align*}
\normalsize
and shall consider the reducible representations of this $SO(3)$ subgroup that are thus provided. For a given $N$, this representation decomposes into the multiplicity-free direct sum of every  $(2\ell+1)$-dimensional irreducible representation of $SO(3)$ with values $\ell=N,\,N-2,\ldots, 1/0$. On the basis vectors \eqref{Basis-Vectors}, the actions \eqref{Actions} take the form
\small
\begin{subequations}
\label{Actions-2}
\begin{align}
a_{1}\kket{m,n}{N}&=\sqrt{m}\,\kket{m-1,n}{N-1},\; &&a_{1}^{\dagger}\kket{m,n}{N}=\sqrt{m+1}\,\kket{m+1,n}{N+1},\\
a_{2}\kket{m,n}{N}&=\sqrt{n}\,\kket{m,n-1}{N-1},\; &&a_{2}^{\dagger}\kket{m,n}{N}=\sqrt{n+1}\,\kket{m,n+1}{N+1},\\
a_{3}\kket{m,n}{N}&=\sqrt{N-m-n}\,\kket{m,n}{N-1},\; &&a_{3}^{\dagger}\kket{m,n}{N}=\sqrt{N-m-n+1}\,\kket{m,n}{N+1}.
\end{align}
\end{subequations}
\normalsize
We use the following notation. Let $B$ be a $3\times 3$ real antisymmetric matrix ($B^{T}=-B$) and $R\in SO(3)$ be the rotation matrix related to $B$ by
\begin{align}
\label{Relation-R-B}
R=e^{B}.
\end{align}
One has of course
\begin{align*}
R^{T}R=RR^{T}=1,
\end{align*}
which in components reads
\begin{align}
\label{Orthogonality}
\sum_{k=1}^{3}R_{ki}R_{kj}=\delta_{ij},\qquad \sum_{k=1}^{3}R_{ik}R_{jk}=\delta_{ij}.
\end{align}
Consider the unitary representation defined by
\begin{align}
\label{Unitary-Rep}
U(R)=\exp\left(\sum_{i,k=1}^{3}B_{ik}a_{i}^{\dagger}a_{k}\right).
\end{align}
The transformations of the generators $a_{i}^{\dagger}$, $a_i$ under the action of $U(R)$ are given by
\begin{align}
\label{Transformation}
U(R)a_{i}^{\dagger}U^{\dagger}(R)=\sum_{k=1}^{3}R_{ki}a_{k}^{\dagger}, \qquad U(R)a_{i}U^{\dagger}(R)=\sum_{k=1}^{3}R_{ki}a_{k}.
\end{align}
Note that $U(R)$ satisfies
\begin{align}
U(RV)=U(R)U(V),\qquad R,V\in SO(3),
\end{align}
as should be for a group representation.
\section{The representation matrix elements as orthogonal polynomials}
In this section, it is shown that the matrix elements of the unitary representations of $SO(3)$ defined in the previous section are expressed in terms of orthogonal polynomials in the two discrete variables $i$, $k$. 

The matrix elements of the unitary operator \eqref{Unitary-Rep} in the basis \eqref{Basis-Vectors} can be cast in the form
\begin{align}
\label{Matrix-Elements}
\BBraket{N}{i,k}{U(R)}{m,n}{N}=W_{i,k;N}P_{m,n}(i,k;N),
\end{align}
where $P_{0,0}(i,k;N)\equiv 1$ and where $W_{i,k;N}$ is defined by
\begin{align}
\label{Vaccuum-Amp}
W_{i,k;N}=\BBraket{N}{i,k}{U(R)}{0,0}{N}.
\end{align}
For notational convenience, we shall drop the explicit dependence of the operator $U$ on the rotation $R$ in what follows.
\subsection{Calculation of the amplitude $W_{i,k;N}$}
An explicit expression can be obtained for the amplitude $W_{i,k;N}$. To that end, one notes using \eqref{Actions-2} that on the one hand
\begin{align*}
\BBraket{N-1}{i,k}{Ua_{1}}{0,0}{N}=0,
\end{align*}
and that on the other hand using \eqref{Transformation} 
\begin{align*}
\BBraket{N-1}{i,k}{Ua_{1}}{0,0}{N}&=\BBraket{N-1}{i,k}{Ua_{1}U^{\dagger}U}{0,0}{N}\\
&=R_{11}\sqrt{i+1}\;\BBraket{N}{i+1,k}{U}{0,0}{N}+R_{21}\sqrt{k+1}\;\BBraket{N}{i,k+1}{U}{0,0}{N}\\
&+R_{31}\sqrt{N-i-k}\;\BBraket{N}{i,k}{U}{0,0}{N}.
\end{align*}
Making use of the definition \eqref{Vaccuum-Amp}, it follows from the above relations that
\begin{align}
\label{Diff-1}
R_{11} \sqrt{i+1}\;W_{i+1,k;N} + R_{21}\sqrt{k+1}\;W_{i,k+1;N} + R_{31}\sqrt{N-i-k} \;W_{i,k;N} =0.
\end{align}
In a similar fashion, starting instead from the relation
\begin{align*}
\BBraket{N-1}{i,k}{Ua_{2}}{0,0}{N}=0,
\end{align*}
one obtains
\begin{align}
\label{Diff-2}
R_{12} \sqrt{i+1}\;W_{i+1,k;N} + R_{22}\sqrt{k+1}\;W_{i,k+1;N} + R_{32}\sqrt{N-i-k}\;W_{i,k;N} =0.
\end{align}
Mindful of \eqref{Orthogonality}, it is easily verified that the common solution to the difference equations \eqref{Diff-1} and \eqref{Diff-2} is
\begin{equation*}
\label{W_C}
W_{i,k;N}= C \: \frac{R_{13}^i R_{23}^k R_{33}^{N-i-k}}{\sqrt{i!k!(N-i-k)!}},
\end{equation*}
where $C$ is an arbitrary constant. This constant can be obtained from the normalization of the basis vectors:
\begin{equation*}
1=\BBraket{N}{0,0}{U^{\dagger}U}{0,0}{N} = \sum_{i+k \leqslant N}\BBraket{N}{0,0}{U^{\dagger}}{i,k}{N}\; \BBraket{N}{i,k}{U}{0,0}{N}=\sum_{i+k \leqslant N} |W_{i,k;N}|^2.
\end{equation*}
Upon using the trinomial theorem, which reads
\begin{equation*}
(z+y+z)^N =\sum_{i+k \le N} \frac{N!}{i!k!(N-i-k)!}\;x^i y^k z^{N-i-k},
\end{equation*}
and the orthogonality relation \eqref{Orthogonality}, one finds that $C = \sqrt{N!}$. Hence the explicit expression for $W_{i,k;N}$ is given by
\begin{equation}
\label{W_C_N}
W_{i,k;N}= R_{13}^i R_{23}^k R_{33}^{N-i-k} \sqrt{\frac{N!}{i!k!(N-i-k)!}}.
\end{equation}
\subsection{Raising relations}
We show that the $P_{m,n}(i,k;N)$ appearing in the matrix elements \eqref{Matrix-Elements} are polynomials of total degree $m+n$ in the variables $i$, $k$  by obtaining raising relations for these polynomials. Consider the matrix element $\BBraket{N}{i,k}{Ua_{1}^{\dagger}}{m,n}{N-1}$. On the one hand, one has
\begin{align*}
\BBraket{N}{i,k}{Ua_{1}^{\dagger}}{m,n}{N-1}=\sqrt{m+1}\,W_{i,k;N}P_{m+1,n}(i,k;N),
\end{align*}
and on the other hand, using \eqref{Transformation}, one finds
\begin{align*}
\BBraket{N}{i,k}{Ua_{1}^{\dagger}}{m,n}{N-1}=\BBraket{N}{i,k}{Ua_{1}^{\dagger}U^{\dagger}U}{m,n}{N-1}=\sum_{m=1}^{3}R_{m,1}\;\BBraket{N}{i,k}{a_{m}^{\dagger}U}{m,n}{N-1}.
\end{align*}
Upon comparing the two preceding expressions and using \eqref{W_C_N}, \eqref{Matrix-Elements}, it follows that
\begin{align}
\nonumber
\sqrt{N(m+1)}&P_{m+1,n}(i,k;N)=\frac{R_{11}}{R_{13}}\,i\,P_{m,n}(i-1,k;N-1)\\
\label{Raising-1}
&+\frac{R_{21}}{R_{23}}\,k\,P_{m,n}(i,k-1;N-1)+\frac{R_{31}}{R_{33}}\,(N-i-k)\,P_{m,n}(i,k;N-1).
\end{align}
In a similar fashion, starting with the matrix element $\BBraket{N}{i,k}{Ua_{2}^{\dagger}}{m,n}{N-1}$, one finds
\begin{align}
\nonumber
\sqrt{N(n+1)}&P_{m,n+1}(i,k;N)=\frac{R_{12}}{R_{13}}\,i\,P_{m,n}(i-1,k;N-1)\\
\label{Raising-2}
&+\frac{R_{22}}{R_{23}}\,k\,P_{m,n}(i,k-1;N-1)+\frac{R_{32}}{R_{33}}\,(N-i-k)\,P_{m,n}(i,k;N-1).
\end{align}
By definition, we have $P_{0,0}(i,k;N) =1$. It is then seen that the formulas \eqref{Raising-1} and \eqref{Raising-2} allow to construct any $P_{m,n}(ik;N)$ from $P_{0,0}(i,k;N)$ through a step by step iterative process and it is observed that these functions are polynomials in the two (discrete) variables $i$, $k$ of total degree $n+m$.
\subsection{Orthogonality relation}
The unitarity of the representation \eqref{Unitary-Rep} can be used to show that the polynomials $P_{m,n}(i,k;N)$ obey an orthogonality relation. Indeed, one has
\begin{align*}
\BBraket{N}{m',n'}{U^{\dagger}U}{m,n}{N}=\sum_{i+k\leqslant N}\BBraket{N}{m'n'}{U^{\dagger}}{i,k}{N}\BBraket{N}{i,k}{U}{m,n}{N}=\delta_{m'm}\delta_{n'n}.
\end{align*}
Upon inserting \eqref{Matrix-Elements}, one finds that the polynomials $P_{m,n}(i,k;N)$ are orthonormal
\begin{align}
\sum_{i+k\leqslant N}w_{i,k;N}P_{m,n}(i,k;N)P_{m',n'}(i,k;N)=\delta_{m'm}\delta_{n'n},
\end{align}
with respect to the discrete weight
\begin{align}
\label{Weight}
w_{i,k;N}=W_{i,k;N}^2=\frac{N!}{i!k!(N-i-k)!}R_{13}^{2i}R_{23}^{2k}R_{33}^{2(N-i-k)}.
\end{align}
\subsection{Lowering relations}
Lowering relations for the polynomials $P_{m,n}(i,k;N)$ can be obtained in a way similar to how the raising relations were found. One first considers the matrix element $\BBraket{N}{i,k}{Ua_1}{m,n}{N+1}$, which leads to the relation
\begin{align}
\nonumber
\sqrt{\frac{m}{N+1}}\,&P_{m-1,n}(i,k;N)=\alpha_1\left[P_{m,n}(i+1,k;N+1)-P_{m,n}(i,k;N+1)\right]\\
\label{Lowering-1}
&+\beta_{1}\left[P_{m,n}(i,k+1;N+1)-P_{m,n}(i,k;N+1)\right].
\end{align}
If one considers instead the matrix element $\BBraket{N}{i,k}{Ua_2}{m,n}{N+1}$, one finds
\begin{align}
\nonumber
\sqrt{\frac{n}{N+1}}\,&P_{m,n-1}(i,k;N)=\alpha_2\left[P_{m,n}(i+1,k;N+1)-P_{m,n}(i,k;N+1)\right]\\
\label{Lowering-2}
&+\beta_{2}\left[P_{m,n}(i,k+1;N+1)-P_{m,n}(i,k;N+1)\right].
\end{align}
In \eqref{Lowering-1} and \eqref{Lowering-2}, the parameters $\alpha$, $\beta$ are given by
\begin{align*}
\alpha_1=R_{11}R_{13},\quad \beta_{1}=R_{21}R_{23},\quad \alpha_2=R_{12}R_{13},\quad \beta_2=R_{22}R_{23}.
\end{align*}
\section{Duality}
A duality relation under the exchange of the variables $i$, $k$ and the degrees $m$, $n$ is obtained in this section for the polynomials $P_{m,n}(i,k;N)$. This property is seen to take a particularly simple form for a set of polynomials $Q_{m,n}(i,k;N)$ which are obtained from $P_{m,n}(i,k;N)$ by a renormalization.

The duality relation for the polynomials $P_{m,n}(i,k;N)$ is obtained by considering the matrix elements $\BBraket{N}{i,k}{U^{\dagger}(R)}{m,n}{N}$ from two different points of view. First one writes
\begin{align}
\label{First-Point}
\BBraket{N}{i,k}{U^{\dagger}(R)}{m,n}{N}=\widetilde{W}_{i,k;N}\widetilde{P}_{m,n;N},
\end{align}
where $\widetilde{P}_{0,0}(i,k;N)=1$ and $\widetilde{W}_{i,k;N}=\BBraket{N}{i,k}{U^{\dagger}}{0,0}{N}$. Since $U^{\dagger}(R)=U(R^{T})$, it follows from \eqref{W_C_N} that
\begin{align*}
\widetilde{W}_{i,k;N}=R_{31}^{i}R_{32}^{k}R_{33}^{N-i-k}\sqrt{\frac{N!}{i!k!(N-i-k)!}},
\end{align*}
and that $\widetilde{P}_{m,n}(i,k;N)$ are the polynomials corresponding to the rotation matrix  $R^{T}$. Second, one instead writes
\begin{align}
\nonumber
\BBraket{N}{i,k}{U^{\dagger}(R)}{m,n}{N}&=\overline{\BBraket{N}{m,n}{U(R)}{i,k}{N}}=\BBraket{N}{m,n}{U(R)}{i,k}{N}\\
\label{Second-Point}
&=W_{m,n;N}P_{i,k}(m,n;N),
\end{align}
where $\overline{x}$ denotes complex conjugation and where the reality of the matrix elements has been used. Upon comparing \eqref{First-Point}, \eqref{Second-Point} and using \eqref{W_C_N},  one directly obtains the duality formula
\begin{align}
\label{Duality-1}
P_{i,k}(m,n;N)=\sqrt{\frac{m!n!(N-m-n)!}{i!k!(-i-k)!}}
\frac{R_{31}^{i}R_{32}^{k}R_{33}^{n+m}}{R_{13}^{m}R_{23}^{n}R_{33}^{i+k}}
\,\widetilde{P}_{m,n}(i,k;N).
\end{align}
It is convenient to introduce the two variable polynomials $Q_{m,n}(i,k;N)$ defined by
\begin{align}
\label{Def-Q}
P_{m,n}(i,k;N)=\sqrt{\frac{N!}{m!n!(N-m-n)!}}\left(\frac{R_{31}}{R_{33}}\right)^{m}\left(\frac{R_{32}}{R_{33}}\right)^{n}Q_{m,n}(i,k;N).
\end{align}
In terms of these polynomials, the duality relation \eqref{Duality-1} reads
\begin{align}
\label{Duality-2}
Q_{i,k}(m,n;N)=\widetilde{Q}_{m,n}(i,k;N),
\end{align}
where the parameters appearing in the polynomial $\widetilde{Q}_{m,n}(i,k;N)$ correspond to the transpose matrix $R^{T}$.
\section{Generating function}
In this section, the generating functions for the multivariate polynomials  $P_{m,n}(i,k;N)$ and $Q_{m,n}(i,k;N)$ are derived using boson calculus. The generating function obtained for $Q_{m,n}(i,k;N)$ is shown to coincide with that of the Rahman polynomials, thus establishing the fact that the polynomials $Q_{m,n}(i,k;N)$ are precisely those defined in \cite{Rahman-2008}. The connection with the parameters used in \cite{Iliev-2012} and \cite{Terwi-2012} is established.

Consider the following generating function for the polynomials $P_{m,n}(i,k;N)$:
\begin{align}
\label{Gen-1}
G(\alpha_1,\alpha_{2},\alpha_{3})=\sum_{m+n+\ell=N}\frac{\alpha_{1}^{m}\alpha_{2}^{n}\alpha_{3}^{\ell}}{\sqrt{m!n!\ell!}}\,W_{i,k;N}P_{m,n}(i,k;N).
\end{align}
Given the definition \eqref{Matrix-Elements} of the matrix elements of $U(R)$, one has
\begin{align*}
G(\alpha_1,\alpha_2,\alpha_3)=\sum_{m+n+\ell=N}\frac{\alpha_{1}^{m}\alpha_{2}^{n}\alpha_{3}^{\ell}}{\sqrt{m!n!\ell!}}\;\BBraket{N}{i,k}{U(R)}{m,n}{N}.
\end{align*}
Upon using \eqref{Ket-Definition} in the above relation, one finds
\begin{align*}
G(\alpha_1,\alpha_2,\alpha_3)=\sum_{m+n+\ell=N}{}_{N}\Braket{i,k}{U\frac{(\alpha_1a_1^{\dagger})^{m}}{m!}\frac{(\alpha_2a_2^{\dagger})^{n}}{n!}\frac{(\alpha_3a_3^{\dagger})^{\ell}}{\ell!}}{0,0,0}.
\end{align*}
Since the rotation operator $U$ preserves any eigenspace with a given energy $N$ and since the states are mutually orthogonal, one can write
\begin{align*}
G(\alpha_1,\alpha_2,\alpha_3)={}_{N}\Braket{i,k}{U\,e^{\alpha_1 a_1^{\dagger}+\alpha_2 a_2^{\dagger}+\alpha_3 a_3^{\dagger}}}{0,0,0}={}_{N}\Braket{i,k}{U\,e^{\alpha_1 a_1^{\dagger}+\alpha_2 a_2^{\dagger}+\alpha_3 a_3^{\dagger}}U^{\dagger}U}{0,0,0}.
\end{align*}
Since $U\ket{0,0,0}=\ket{0,0,0}$ and 
\begin{align*}
Ue^{\sum_{s}\alpha_{s}a_{s}^{\dagger}}U^{\dagger}=e^{\sum_{s}\alpha_{s}Ua_{s}^{\dagger}U^{\dagger}}=e^{\sum_{s t}\alpha_{s}R_{t s}a_{t}^{\dagger}}=e^{\sum_{t}\beta_{t}a_{t}^{\dagger}},
\end{align*}
where $$\beta_t=\sum_{s=1}^{3} R_{ts}\alpha_{s},$$ it follows that 
\begin{align*}
G(\alpha_1,\alpha_2,\alpha_3)={}_{N}\Braket{i,k}{e^{\beta_1 a_{1}^{\dagger}+\beta_2 a_2^{\dagger}+\beta_{3}a_{3}^{\dagger}}}{0,0,0}=\sum_{p,q,r}\frac{\beta_1^{p}\beta_{2}^{q}\beta_{3}^{r}}{\sqrt{p!q!r!}}\braket{i,k,j}{p,q,r},
\end{align*}
with $j=N-i-k$. Using the orthogonality of the basis states, one thus obtains
\begin{align}
\label{Gen-2}
G(\alpha_1,\alpha_2,\alpha_3)=\frac{\beta_1^{i}\beta_{2}^{k}\beta_{3}^{N-i-k}}{\sqrt{i!k!(N-i-k)!}}.
\end{align}
Upon comparing the expressions \eqref{Gen-1}, \eqref{Gen-2} and recalling the expression for $W_{i,k;N}$ given by \eqref{W_C_N}, one finds
\begin{align}
\label{Gen-3}
\frac{\beta_1^{i}\beta_2^{k}\beta_{3}^{N-i-k}}{\sqrt{N!}}=R_{13}^{i}R_{23}^{k}R_{33}^{N-i-k}\sum_{m+n\leqslant N}\frac{\alpha_1^{m}\alpha_{2}^{n}\alpha_{3}^{N-n-m}}{\sqrt{m!n!(N-n-m)!}}\;P_{m,n}(i,k;N).
\end{align}
Taking $\alpha_1=u$, $\alpha_2=v$ and $\alpha_3=1$ in  \eqref{Gen-3}, one obtains the following generating function for the polynomials $P_{m,n}(i,k;N)$:
\small
\begin{align}
\nonumber
\left(1+\frac{R_{11}}{R_{13}}u+\frac{R_{12}}{R_{13}}v\right)^{i}
&\left(1+\frac{R_{21}}{R_{23}}u+\frac{R_{22}}{R_{23}}v\right)^{k}
\left(1+\frac{R_{31}}{R_{33}}u+\frac{R_{32}}{R_{33}}v\right)^{N-i-k}\\
&=\sum_{m+n\leqslant N}\sqrt{\frac{N!}{m!n!(N-m-n)!}}\;P_{m,n}(i,k;N)\;u^{m}v^{n}.
\end{align}
\normalsize
In terms of the polynomials $Q_{m,n}(i,k;N)$, the relation \eqref{Gen-3} reads
\small
\begin{align*}
&\left(\frac{R_{11}}{R_{13}}\alpha_1+\frac{R_{12}}{R_{13}}\alpha_2+\alpha_3\right)^{i}
\left(\frac{R_{21}}{R_{23}}\alpha_1+\frac{R_{22}}{R_{23}}\alpha_2+\alpha_3\right)^{k}
\left(\frac{R_{31}}{R_{33}}\alpha_1+\frac{R_{32}}{R_{33}}\alpha_2+\alpha_3\right)^{N-i-k}\\
&=\sum_{m+n\leqslant N}\frac{N!}{m!n!(N-m-n)!}\;\left(\frac{R_{31}}{R_{33}}\right)^{m}\left(\frac{R_{32}}{R_{33}}\right)^{n} Q_{m,n}(i,k;N)\;\alpha_{1}^{m}\alpha_{2}^{n}\alpha_{3}^{N-m-n}.
\end{align*}
\normalsize
Upon taking instead
\begin{align*}
\alpha_1=\frac{R_{33}}{R_{31}}z_1,\quad \alpha_{2}=\frac{R_{33}}{R_{32}}z_2,\quad \alpha_3=1,
\end{align*}
one finds the following generating function for the polynomials $Q_{m,n}(i,k;N)$:
\small
\begin{align}
\nonumber
\left(1+\frac{R_{11}R_{33}}{R_{13}R_{31}}z_1+\frac{R_{12}R_{33}}{R_{13}R_{32}}z_2\right)^{i}
&\left(1+\frac{R_{21}R_{33}}{R_{23}R_{31}}z_1+\frac{R_{22}R_{33}}{R_{23}R_{32}}z_2\right)^{k}
\left(1+z_1+z_2\right)^{N-i-k}\\
&=\sum_{m+n\leqslant N}\frac{N!}{m!n!(N-m-n)!}\; Q_{m,n}(i,k;N)\;z_{1}^{m}z_{2}^{n}.
\label{Gen-Q}
\end{align}
\normalsize
The formula \eqref{Gen-Q} lends itself to comparison with the generating function which can be taken to define the bivariate Krawtchouk polynomials (see \eqref{Annex-Generating}). Up to an obvious identification of the indices $(i,k)\rightarrow (\widetilde{m}_1,\widetilde{m}_2)$ and $(m,n)\rightarrow (m_1,m_2)$, the generating function \eqref{Gen-Q} coincides with \eqref{Annex-Generating} and shows that the polynomials $Q_{m,n}(i,k;N)$ are the same as the Rahman polynomials $Q(m,\widetilde{m})$ if one takes
\begin{subequations}
\label{Para-1}
\begin{align}
&u_{11}=\frac{R_{11}R_{33}}{R_{13}R_{31}},\quad u_{12}=\frac{R_{12}R_{33}}{R_{13}R_{32}},\\
&u_{21}=\frac{R_{21}R_{33}}{R_{23}R_{31}},\quad u_{22}=\frac{R_{22}R_{33}}{R_{23}R_{32}}.
\end{align}
\end{subequations}
The parametrization \eqref{Para-1} can be related to the one in terms of the four parameters $p_1$, $p_2$, $p_3$ and $p_4$ that is customarily used to define the Rahman polynomials. The reader is referred to the Appendix for the precise correspondence and for further details concerning the parametrizations.
\section{Recurrence relations and difference equations}
In this section, the algebraic setting is used to derive the recurrence relations and difference equations that the polynomials $P_{m,n}(i,k;N)$ and $Q_{m,n}(i,k;N)$ satisfy. We note that these have been obtained previously in \cite{Grunbaum-2007} and \cite{Terwi-2012} (see also \cite{Xu-2004}). It is interesting to see how easily the recurrence relations and difference equations follow from the group-theoretic interpretation. See also \cite{Area-2012,Iliev-2007,Xu-2004}.
\subsection{Recurrence relations}
The obtain the recurrence relations for the polynomials $P_{m,n}(i,k;N)$, one begins by considering the matrix element $\BBraket{N}{i,k}{a_{1}^{\dagger}a_1U}{m,n}{N}$. One has on the one hand
\begin{align*}
\BBraket{N}{i,k}{a_{1}^{\dagger}a_1U}{m,n}{N}=i\;\BBraket{N}{i,k}{U}{m,n}{N},
\end{align*}
and on the other hand
\begin{align*}
\nonumber
\BBraket{N}{i,k}{a_{1}^{\dagger}a_1U}{m,n}{N}&=\BBraket{N}{i,k}{UU^{\dagger}a_{1}^{\dagger}a_1U}{m,n}{N}\\
&=\sum_{r,s=1}^{3}R_{r,1}R_{s,1}\;\BBraket{N}{i,k}{Ua_{r}^{\dagger}a_{s}}{m,n}{N}.
\end{align*}
Upon comparing the above equations and using \eqref{Matrix-Elements} and \eqref{W_C_N}, one finds
\small
\begin{align}
\nonumber
i&\,P_{m,n}(i,k;N)=\left[R_{11}^2 m + R_{12}^2 n + R_{13}^2 (N-m-n) \right]\, P_{m,n}(i,k;N) \\
\label{Recurrence-P-1}
&+
R_{11} R_{12} \left[\sqrt{m(n+1)}\,P_{m-1,n+1}(i,k;N)+\sqrt{n(m+1)}\,P_{m+1,n- 1}(i,k;N) \right] \\
\nonumber
&+
R_{11} R_{13} \left[\sqrt{m(N-m-n+1)}\,P_{m-1,n}(ik;N)+\sqrt{(m+1)(N-m-n)}\,P_{m+1,n}(i,k;N)\right] \\
\nonumber
&+
R_{12} R_{13} \left[\sqrt{n(N-m-n+1)}  P_{m,n-1}(ik;N)   + \sqrt{(n+1)(N-m-n)} P_{m,n+1}(ik;N)    \right].
\end{align}
\normalsize
Proceeding similarly from the matrix element $\BBraket{N}{i,k}{a_{2}^{\dagger}a_{2}U}{m,n}{N}$, one obtains
\small
\begin{align}
\nonumber
k\,&P_{m,n}(i,k;N)=
\left[R_{21}^2 m + R_{22}^2 n + R_{23}^2 (N-m-n) \right]\,P_{m,n}(i,k;N)
\\
\label{Recurrence-P-2}
&+R_{21} R_{22} \left[\sqrt{m(n+1)}\,P_{m-1,n+1}(i,k;N)  + \sqrt{n(m+1)}\,P_{m+1,n- 1}(i,k;N)\right]
\\
\nonumber
&+R_{21} R_{23} \left[\sqrt{m(N-m-n+1)}\,P_{m-1,n}(i,k;N)+\sqrt{(m+1)(N-m-n)} P_{m+1,n}(i,k;N)\right]
\\
\nonumber
&+
R_{22} R_{23}\left[\sqrt{n(N-m-n+1)}\,P_{m,n-1}(i,k;N)+\sqrt{(n+1)(N-m-n)}\,P_{m,n+1}(i,k;N)\right].
\end{align}
\normalsize
In terms of the polynomials $Q_{m,n}(i,k;N)$ defined by \eqref{Def-Q}, the recurrence relations \eqref{Recurrence-P-1} and \eqref{Recurrence-P-2} become
\begin{align} 
\nonumber
&i\,Q_{m,n}(i,k;N)=\left[R_{11}^2\,m+R_{12}^2\,n + R_{13}^2\,(N-m-n) \right]\,Q_{m,n}(i,k;N) \\
\label{Recurrence-Q-1}
&+
\frac{R_{11} R_{12} R_{32}}{R_{31}}\,m\,Q_{m-1,n+1}(i,k;N)+\frac{R_{11} R_{12} R_{31}}{R_{32}}\,n\,Q_{m+1,n-1}(i,k;N)
\\
\nonumber
&+
\frac{R_{11} R_{13} R_{33}}{R_{31}}\,m\,Q_{m-1,n}(i,k;N)+\frac{R_{11} R_{13} R_{31}}{R_{33}}\,(N-m-n)\,Q_{m+1,n}(i,k;N)
\\
\nonumber
&+
\frac{R_{12} R_{13} R_{33}}{R_{32}}\,n\,Q_{m,n-1}(i,k;N)+\frac{R_{12} R_{13} R_{32}}{R_{33}}\,(N-m-n)\,Q_{m,n+1}(i,k;N),
\end{align}
and
\begin{align} 
\nonumber
&k\,Q_{m,n}(i,k;N)=\left[R_{21}^2\,m+R_{22}^2\,n + R_{23}^2\,(N-m-n) \right]\,Q_{m,n}(i,k;N) \\
\label{Recurrence-Q-2}
&+
\frac{R_{21} R_{22} R_{32}}{R_{31}}\,m\,Q_{m-1,n+1}(i,k;N)+\frac{R_{21} R_{22} R_{31}}{R_{32}}\,n\,Q_{m+1,n-1}(i,k;N)
\\
\nonumber
&+
\frac{R_{21} R_{23} R_{33}}{R_{31}}\,m\,Q_{m-1,n}(i,k;N)+\frac{R_{21} R_{23} R_{31}}{R_{33}}\,(N-m-n)\,Q_{m+1,n}(i,k;N)
\\
\nonumber
&+
\frac{R_{22} R_{23} R_{33}}{R_{32}}\,n\,Q_{m,n-1}(i,k;N)+\frac{R_{22} R_{23} R_{32}}{R_{33}}\,(N-m-n)\,Q_{m,n+1}(i,k;N).
\end{align}
\subsection{Difference equations}
To obtain the difference equations satisfied by the polynomials $Q_{m,n}(i,k;N)$, one could consider the matrix elements $\BBraket{N}{i,k}{Ua_{j}^{\dagger}a_{j}}{m,n}{N}$, $j=1,2$ and proceed along the same lines as for the recurrence relations. It is however easier to proceed directly from the recurrence relations \eqref{Recurrence-Q-1}, \eqref{Recurrence-Q-2} and use the duality relation \eqref{Duality-2}. To illustrate the procedure, consider the left hand side of \eqref{Recurrence-Q-1}. Using the duality \eqref{Duality-2}, one may write
\begin{align*}
i Q_{m,n}(i,k;N)=i\,\widetilde{Q}_{i,k}(m,n;N)=m\,\widetilde{Q}_{m,n}(i,k;N),
\end{align*}
where in the last step the replacements $m\leftrightarrow i$ and $n\leftrightarrow k$ were performed. Since $\widetilde{Q}_{m,n}(i,k;N)$ is obtained from $Q_{m,n}(i,k;N)$ by replacing the rotation parameters by the elements of the transposed, it is seen that the recurrence relations \eqref{Recurrence-Q-1} and \eqref{Recurrence-Q-2} can be turned into difference equations by operating the substitutions $m\leftrightarrow i$, $n\leftrightarrow k$ and replacing the parameters of $R$ by those of $R^{T}$. Applying this procedure, it then follows easily that
\begin{align} 
\nonumber
&m\,Q_{m,n}(i,k;N)=\left[R_{11}^2\,i+R_{21}^2\,k + R_{31}^2\,(N-i-k) \right]\,Q_{m,n}(i,k;N) \\
\label{Difference-Q-1}
&+
\frac{R_{11} R_{21} R_{23}}{R_{13}}\,i\,Q_{m,n}(i-1,k+1;N)+\frac{R_{11} R_{21} R_{13}}{R_{23}}\,k\,Q_{m,n}(i+1,k-1;N)
\\
\nonumber
&+
\frac{R_{11} R_{31} R_{33}}{R_{13}}\,i\,Q_{m,n}(i-1,k;N)+\frac{R_{11} R_{31} R_{13}}{R_{33}}\,(N-i-k)\,Q_{m,n}(i+1,k;N)
\\
\nonumber
&+
\frac{R_{21} R_{31} R_{33}}{R_{23}}\,k\,Q_{m,n}(i,k-1;N)+\frac{R_{21} R_{31} R_{23}}{R_{33}}\,(N-i-k)\,Q_{m,n}(i,k+1;N),
\end{align}
and 
\begin{align} 
\nonumber
&n\,Q_{m,n}(i,k;N)=\left[R_{12}^2\,i+R_{22}^2\,k + R_{32}^2\,(N-i-k) \right]\,Q_{m,n}(i,k;N) \\
\label{Difference-Q-2}
&+
\frac{R_{12} R_{22} R_{23}}{R_{13}}\,i\,Q_{m,n}(i-1,k+1;N)+\frac{R_{12} R_{22} R_{13}}{R_{23}}\,k\,Q_{m,n}(i+1,k-1;N)
\\
\nonumber
&+
\frac{R_{12} R_{32} R_{33}}{R_{13}}\,i\,Q_{m,n}(i-1,k;N)+\frac{R_{12} R_{32} R_{13}}{R_{33}}\,(N-i-k)\,Q_{m,n}(i+1,k;N)
\\
\nonumber
&+
\frac{R_{22} R_{32} R_{33}}{R_{23}}\,k\,Q_{m,n}(i,k-1;N)+\frac{R_{22} R_{32} R_{23}}{R_{33}}\,(N-i-k)\,Q_{m,n}(i,k+1;N),
\end{align}
Similar formulas can be obtained straightforwardly for the polynomials $P_{m,n}(i,k;N)$.
\section{Integral representation}
In this section, a relation between the Hermite polynomials and the bivariate Krawtchouk polynomials $P_{m,n}(i,k;N)$ is found. This relation allows for the presentation of an integral formula for these polynomials. To obtain these formulas, one begins by considering the matrix element $$\langle x_1,x_2,x_3\rvert U(R)\ket{m,n}_{N},$$ from two points of view. By acting with $U(R)$ on the vector $\kket{m,n}{N}$ and using \eqref{Matrix-Elements} and \eqref{Coordinate-Rep}, one finds
\begin{align*}
\langle x_1,x_2,x_3\rvert U(R)\ket{m,n}_{N}&=\sqrt{\frac{N!}{2^{N}\pi^{3/2}}}\,e^{-(x_1^2+x_2^2+x_3^2)/2}\,\sum_{i+k\leqslant N}\frac{R_{13}^{i}R_{23}^{k}R_{33}^{N-i-k}}{i!k!(N-i-k)!}P_{m,n}(i,k;N)\\
&\times H_{i}(x_1)H_{k}(x_2)H_{N-i-k}(x_3).
\end{align*}
By acting with $U^{\dagger}(R)$ on $\bra{x_1,x_2,x_3}$ and using \eqref{Coordinate-Rep}, one finds
\begin{align*}
\langle x_1,x_2,x_3\rvert U(R)\ket{m,n}_{N}&=\frac{e^{-(\widetilde{x}_1^2+\widetilde{x}_2^2+\widetilde{x}_3^2)/2}}{\sqrt{2^{N}\pi^{3/2}m!n!(N-m-n)!}}\,H_{m}(\widetilde{x}_1)H_{n}(\widetilde{x}_2)H_{N-m-n}(\widetilde{x}_3),
\end{align*}
where $(\widetilde{x}_1,\widetilde{x}_2,\widetilde{x}_3)^{T}=R^{T}(x_1,x_2,x_3)^{T}$.
Since obviously $x_1^2+x_2^2+x_3^2=\widetilde{x}_1^2+\widetilde{x}_2^2+\widetilde{x}_3^2$, one finds from the above
\small
\begin{align}
\nonumber
&\sqrt{\frac{1}{N!m!n!(N-m-n)!}}\,H_{m}(\widetilde{x}_1)H_{n}(\widetilde{x}_2)H_{N-m-n}(\widetilde{x}_3)\\
&=\sum_{i+k\leqslant N}\frac{R_{13}^{i}R_{23}^kR_{33}^{N-i-k}}{i!k!(N-i-k)!}\,P_{m,n}(i,k;N)H_{i}(x_1)H_{k}(x_2)H_{N-i-k}(x_3).
\label{P-VS-Hermite}
\end{align}
\normalsize
Using the relation \eqref{P-VS-Hermite} and the well-known orthogonality relation satisfied by the Hermite polynomials \cite{Koekoek-2010}, one obtains the following integral representation for the polynomials $P_{m,n}(i,k;N)$:
\small
\begin{align}
\nonumber
&P_{m,n}(i,k;N)=\frac{R_{13}^{-i}R_{23}^{-k}R_{33}^{i+k-N}}{2^{N}\pi^{3/2}}\sqrt{\frac{1}{N!m!n!(N-m-n)!}}\\
&\times \int_{\mathbb{R}^{3}}e^{-(x_1^2+x_2^2+x_3^2)}H_{m}(\widetilde{x}_1)H_{n}(\widetilde{x}_2)H_{N-m-n}(\widetilde{x}_3)H_{i}(x_1)H_{k}(x_2)H_{N-i-k}(x_3)\;dx_1dx_2dx_3.
\end{align}
\normalsize

\section{Rotations in coordinate planes and univariate Krawtchouk polynomials}
It has been assumed so far generically that the entries $R_{ik}$, $i,k=1,2,3$, of the rotation matrix $R$ are non-zero. We shall now consider the degenerate cases corresponding to when rotations are restricted to coordinate planes and when the matrix $R$ has thus four zero entries. We shall confirm, as expected, that the representation matrix elements $\BBraket{N}{i,k}{U(R)}{m,n}{N}$ are then expressed in terms of univariate Krawtchouk polynomials. With $J$ a non-negative integer, the one-dimensional Krawtchouk polynomials $k_{n}(x;p;J)$ that we shall use are defined by
\begin{align}
\label{Def-Kr}
k_{n}(x;p;J)=(-J)_{n} \;\pFq{2}{1}{-n,-x}{-J}{\frac{1}{p}}=(-J)_{n}\sum_{k=0}^{\infty}\frac{(-n)_{k}(-x)_{k}}{k!(-J)_{k}}\left(\frac{1}{p}\right)^{k},
\end{align}
where
\begin{align*}
(a)_{k}=a(a+1)\cdots(a+k-1),\qquad k\geqslant 1, \quad (a)_0=1.
\end{align*}
These polynomials are orthogonal with respect to the binomial distribution and satisfy the orthogonality relation
\begin{align*}
\sum_{x=0}^{J}\frac{J!}{x!(J-x)!}p^{x}(1-p)^{J-x}k_{m}(x;p;J)k_{n}(x;p;J)=(-1)^{n}n!(-J)_{n}\left[\frac{(1-p)}{p}\right]^{n}\delta_{mn}.
\end{align*}
They are related as follows to the monic polynomials $q_{n}(x)$:
\begin{align*}
q_{n}(x)=p^{n}k_n(x;p,J).
\end{align*}
which satisfy the following three-term recurrence relation \cite{Koekoek-2010}:
\begin{align}
\label{Recu-Kr}
x q_{n}(x)=q_{n+1}(x)+[p(J-n)+n(1-p)]q_{n}(x)+np(1-p)(J+1-n)q_{n-1}(x).
\end{align}
Consider the clockwise rotation $R_{(yz)}(\theta)$ by an angle $\theta$ in the $(yz)$ plane and the clockwise rotation $R_{(xz)}(\chi)$ by an angle $\chi$ in the $(xz)$ plane. They correspond to the matrices
\begin{align}
\label{Rotations}
R_{(yz)}(\theta)=
\begin{pmatrix}
1 & 0 & 0\\
0 & \cos \theta & \sin \theta\\
0 & -\sin \theta & \cos \theta
\end{pmatrix},
\quad
R_{(xz)}(\chi)=
\begin{pmatrix}
\cos \chi & 0 & -\sin\chi\\
0 & 1 & 0\\
\sin\chi & 0 & \cos \chi
\end{pmatrix}.
\end{align}
Note that their product 
\begin{align}
\label{Rotation-Tratnik}
R_{(yz)}(\theta)R_{(xz)}(\chi)=
\begin{pmatrix}
\cos \chi & 0 & -\sin\chi\\
\sin\theta\sin\chi & \cos\theta & \sin\theta\cos\chi\\
\cos\theta\sin\chi & -\sin\theta & \cos\theta\cos\chi
\end{pmatrix},
\end{align}
has one zero entry ($R_{12}=0$). The rotations $R_{(yz)}(\theta)$ and $R_{(xz)}(\chi)$ are unitarily represented by the operators
\begin{align*}
U_{(yz)}(\theta)=e^{\theta(a_{2}^{\dagger}a_{3}-a_{3}^{\dagger}a_{2})},\quad \text{and}\quad U_{(xz)}(\chi)=e^{\chi(a_{3}^{\dagger}a_{1}-a_{1}^{\dagger}a_{3})},
\end{align*}
respectively. We now wish to obtain the matrix elements $\BBraket{N}{i,k}{U_{(yz)(\theta)}}{m,n}{N}$ and $\BBraket{N}{i,k}{U_{(xz)}(\chi)}{m,n}{N}$ of these operators. This can be done by adopting the same approach as in Section 3. Details shall be given for the rotation about the $x$ axis. Since $U_{(yz)}(\theta)$ leaves $a_{1}$ and $a_{1}^{\dagger}$ unchanged and acts trivially on the first quantum number, it is readily seen that
\begin{align}
\label{1}
\BBraket{N}{i,k}{U_{(yz)}(\theta)}{m,n}{N}=\delta_{im}\;\BBraket{J}{k}{U_{(yz)}(\theta)}{n}{J},
\end{align}
where
$$
\kket{\ell}{J}\equiv \kket{i,\ell}{N},
$$
and with $J=N-i$ and $\ell$ taking the values $0,1,\ldots,J$. Given that 
\begin{align*}
U^{\dagger}_{(yz)}(\theta)a_{2}U_{(yz)}(\theta)=a_2\cos \theta+a_3\sin \theta,
\end{align*}
and that $a_{2}^{\dagger}$ transforms in the same way, the identity
\begin{align*}
\BBraket{J}{k}{a_{2}^{\dagger}a_{2}U_{(yz)}(\theta)}{n}{J}=\BBraket{J}{K}{U_{(yz)}(\theta)U_{(yz)}^{\dagger}(\theta)a_{2}^{\dagger}a_{2}U_{(yz)}(\theta)}{n}{J},
\end{align*}
yields the recurrence relation
\begin{align}
\nonumber
k\;\BBraket{J}{k}{U_{(yz)}(\theta)}{n}{J}&=\left[n\cos^2\theta+(J-n)\sin^2\theta\right]\BBraket{J}{k}{U_{(yz)}(\theta)}{n}{J}
\\
\nonumber
&+\cos\theta\sin\theta\Big[\sqrt{(n+1)(J-n)}\,\BBraket{J}{k}{U_{(yz)}(\theta)}{n+1}{J}
\\
\nonumber
&+\sqrt{n(J-n+1)}\;\BBraket{J}{k}{U_{(yz)}(\theta)}{n-1}{J}\Big].
\end{align}
Now write 
\begin{align}
\label{2}
\BBraket{J}{k}{U_{(yz)}(\theta)}{n}{J}=\BBraket{J}{k}{U_{(yz)}(\theta)}{0}{J}\;\sqrt{\frac{(-1)^{n}}{n!(-J)_{n}}}\,\frac{q_{n}(k)}{\cos^{n}\theta\sin^{n}\theta},
\end{align}
to find that indeed $q_{n}(k)$ verifies the three-term recurrence relation \eqref{Recu-Kr} of the monic Krawtchouk polynomials with $p=\sin^2\theta$.
Using the identity
\begin{align*}
\BBraket{J-1}{k}{U_{(yz)}(\theta)a_2 U_{(yz)}^{\dagger}(\theta)U_{(yz)}(\theta)}{0}{J}=
\BBraket{J-1}{k}{U_{(yz)}(\theta)a_2}{n}{J}=0,
\end{align*}
we find the prefactor to obey the two-term recurrence relation
\begin{align*}
\sqrt{k+1}\,\cos\theta \,\BBraket{J}{k+1}{U_{(yz)}(\theta)}{0}{J}=\sqrt{J-k}\,\sin\theta \,\BBraket{J}{k}{U_{(yz)}(\theta)}{0}{J},
\end{align*}
which has for solution
\begin{align}
\label{3}
\BBraket{J}{k}{U_{(yz)}(\theta)}{0}{J}=\BBraket{J}{0}{U_{(yz)}(\theta)}{0}{J}\,\sqrt{\frac{J!}{k!(J-k)!}}\tan^{k}\theta.
\end{align}
The ground state expectation value is again found from the normalization of the state vectors. One has
\begin{align*}
\nonumber
1&=\BBraket{J}{0}{U_{(yz)}(\theta)U_{(yz)}^{\dagger}(\theta)}{0}{J}=\sum_{k=0}^{J}\BBraket{J}{0}{U_{(yz)}(\theta)}{k}{J}\BBraket{J}{k}{U_{(yz)}(\theta)}{0}{J}\\
&=|\BBraket{J}{0}{U_{(yz)}(\theta)}{0}{J}|^2\;\sum_{k=0}^{J}\frac{J!}{k!(J-k)!}\tan^{k}\theta,
\end{align*}
which gives
\begin{align}
\label{4}
\BBraket{J}{0}{U_{(yz)}(\theta)}{0}{J}=\cos^{J}\theta.
\end{align}
Putting \eqref{1}, \eqref{2}, \eqref{3} and \eqref{4} together, one finds
\begin{align}
\nonumber
&\BBraket{N}{i,k}{U_{(yz)}(\theta)}{m,n}{N}=\delta_{im}
\\
\label{Matrix-Elements-Rx}
&\times\sqrt{\frac{(-1)^{n}(N-i)!}{k!n!(N-i-k)!(i-N)_{n}}}\cos^{N-i}\theta \tan^{k+n}\theta\,k_{n}(k;\sin^2\theta;N-i).
\end{align}
The matrix elements of $U_{(xz)}(\chi)$ can be obtained in an identical fashion and one finds
\begin{align}
\nonumber
&\BBraket{N}{i,k}{U_{(xz)}(\chi)}{m,n}{N}=\delta_{kn}
\\
\label{Matrix-Elements-Ry}
&\times(-1)^{i+m}\sqrt{\frac{(-1)^{m}(N-n)!}{i!m!(N-n-i)!(n-N)_{m}}}\cos^{N-n}\chi \tan^{i+m}\chi\;k_{m}(i;\sin^2\chi;N-n).
\end{align}
Note that in this case, the Kronecker delta involves the second quantum numbers as those are the ones that are left unscathed by the rotation about the $y$ axis.
\section{The bivariate Krawtchouk-Tratnik as special cases}
In this section, the recurrence relations derived in Section 6 will be used to show that the Krawtchouk-Tratnik polynomials are specializations of the Rahman or Krawtchouk-Griffiths polynomials. Aspects of the relation between the two sets of polynomials are also discussed in \cite{Miki-2013}.

The bivariate Krawtchouk-Tratnik polynomials, denoted $K_2(m,n;i,k;\mathfrak{p}_1,\mathfrak{p}_2;N)$, are a family of  polynomials introduced by Tratnik in \cite{Tratnik-1991}. These polynomials are defined in terms of the univariate Krawtchouk polynomials as follows:
\begin{align}
\label{Def-Tratnik}
K_2(m,n;i,k;\mathfrak{p}_1,\mathfrak{p}_2;N)=\frac{1}{(-N)_{n+m}}
k_{m}(i;\mathfrak{p_1};N-n)k_{n}(k;\frac{\mathfrak{p_2}}{1-\mathfrak{p}_1};N-i),
\end{align}
where $k_{n}(x;p,J)$ is as in \eqref{Def-Kr}. They are orthogonal with respect to the trinomial distribution
\begin{align*}
w_{ik}=\frac{N!}{i!k!(N-i-k)!}\mathfrak{p}_1^{i}\mathfrak{p}_2^{k}(1-\mathfrak{p}_1-\mathfrak{p}_2)^{N-i-k}.
\end{align*}
Their bispectral properties have been determined by Geronimo and Iliev in \cite{Iliev-2010} to which the reader is referred (see the Appendix) for details.

Consider at this point the recurrence relations \eqref{Recurrence-Q-1}, \eqref{Recurrence-Q-2} and take $R_{12}=0$. It is observed that in this case the recurrence relation \eqref{Recurrence-Q-1} simplifies considerably. Indeed, one has
\begin{align} 
\nonumber
i\,&Q_{m,n}(i,k;N)=\left[R_{11}^2 m + R_{31}^2 (N-m-n) \right] Q_{m,n}(i,k;N)
\\
\label{Recurrence-Q-1-Tratnik}
&+
\frac{R_{11}R_{31}R_{33}}{R_{13}}\;m\;Q_{m-1,n}(i,k;N)+\frac{R_{11} R_{31} R_{13}}{R_{33}}\,(N-n-m)\,Q_{m+1,n}(i,k;N). 
\end{align} 
The condition $R_{12}=0$ implies that the matrix elements now verify the orthogonality relation $R_{11}R_{31}+R_{13}R_{33}=0$; hence \eqref{Recurrence-Q-1-Tratnik} reduces to
\begin{align}
\nonumber
i\,Q_{m,n}(i,k;N) &=R_{11}^2\,m\,\left[Q_{m,n}(i,k;N)-Q_{m-1,n}(i,k;N) \right]\\
& + R_{13}^2\,(N-m-n)\,\left[Q_{m,n}(i,k;N)-Q_{m+1,n}(i,k;N)\right].
\label{Recurrence-Q-1-Tratnik-2}
\end{align}  
Upon comparing the formula \eqref{Recurrence-Q-1-Tratnik-2} and the formula of Geronimo and Iliev \eqref{Tratnik-A}, it is seen that they coincide provided that
\begin{align}
\label{Para-Tratnik}
\mathfrak{p}_1=R_{13}^2,\quad \mathfrak{p}_2=R_{23}^2.
\end{align}
One also checks (see Appendix A for more details) that the second relation \eqref{Annex-Recu-2} is recovered under this identification of the parameters. This shows that the Krawtchouk-Tratnik polynomials are a special case of the general bivariate Krawtchouk polynomials $Q_{m,n}(i,k;N)$ where the rotation matrix has one of its entries equal to zero, namely $R_{12}=0$. The Krawtchouk-Tratnik polynomials thus arise when the rotation matrix can be written as the composition of two rotations: one in the plane $(yz)$ and the other in the plane $(xz)$, this explains why the Krawtchouk-Tratnik polynomials only depend on two parameters. In the next section, the addition formulas for the general polynomials $P_{m,n}(i,k;N)$ provided by the group product will give a direct derivation of the Tratnik formula \eqref{Def-Tratnik}.

\section{Addition formulas}
In this section, the group product is used to derive an addition formula for the general bivariate Krawtchouk polynomials $P_{m,n}(i,k;N)$. In the special case where the rotation is a product of plane rotations around the $x$ and $y$ axes, the addition formula is used to recover the explicit expression of the Krawtchouk-Tratnik polynomials. In the general case, in which the rotation is a product of three rotations, the addition formula yields an expansion formula of the general polynomials $Q_{m,n}(i,k;N)$ in terms of the Krawtchouk-Tratnik polynomials.

\subsection{General addition formula}
Let $A$ and $B$ be two arbitrary rotation matrices. Their product $C=AB$ is also a rotation matrix. Denote by $U(C)$, $U(A)$ and $U(B)$ the unitary operators representing the rotations $C$, $A$ and $B$ as specified by \eqref{Unitary-Rep}. For a given $N$, to each of these rotations is associated a system of bivariate Krawtchouk polynomials that can be designated by $P_{m,n}^{(C)}(i,k;N)$, $P_{m,n}^{(A)}(i,k;N)$ and $P_{m,n}^{(B)}(i,k;N)$. Since $U$ defines a representation, one has $U(C)=U(A)U(B)$ and hence it follows that
\begin{align}
\label{Addition-Matrix-Elements}
\BBraket{N}{i,k}{U(C)}{m,n}{N}=\sum_{q+r\leqslant N}\BBraket{N}{i,k}{U(A)}{q,r}{N}\;\BBraket{N}{q,r}{U(B)}{m,n}{N}.
\end{align}
In terms of the polynomials, this identity amounts to the general addition formula
\begin{align}
\label{Addition-Formula}
\left(\frac{W_{i,k;N}^{(C)}}{W_{i,k;N}^{(A)}}\right)P_{m,n}^{(C)}(i,k;N)=\sum_{q+r\leqslant N}W_{q,r;N}^{(B)}P_{q,r}^{(A)}(i,k;N)P_{m,n}^{(B)}(q,r;N).
\end{align}
\subsection{The Tratnik expression}
The addition property \eqref{Addition-Matrix-Elements} of the matrix elements can be used to recover the explicit expression for the bivariate Krawtchouk-Tratnik polynomials. It has already been established that the general polynomials $Q_{m,n}(i,k;N)$ correspond to the Tratnik ones when $R_{12}=0$. We saw that this occurs when the rotation is of the form $C=AB$ with $A=R_{(yz)}(\theta)$ and $B=R_{(xz)}(\chi)$. Considering the left hand side of \eqref{Addition-Matrix-Elements} and using \eqref{Def-Q}, it follows that
\begin{align}
\nonumber
&\BBraket{N}{i,k}{U(C(\theta,\chi))}{m,n}{N}=\frac{N!\,C_{33}^{N}}{\sqrt{i!k!(N-i-k)!m!n!(N-m-n)!}}
\\
\label{First}
&\times\left(\frac{C_{13}}{C_{33}}\right)^{i}
\left(\frac{C_{23}}{C_{33}}\right)^{k}
\left(\frac{C_{31}}{C_{33}}\right)^{m}
\left(\frac{C_{32}}{C_{33}}\right)^{n} K_2(m,n;i,k;\sin^2\chi,\sin^2\theta\cos^2\chi;N),
\end{align}
\normalsize
where the parameter identification \eqref{Para-Tratnik} has been used and where the corresponding rotation matrix $C(\theta,\chi)$ is given in \eqref{Rotation-Tratnik}. Considering the right hand side of \eqref{Addition-Matrix-Elements} and recalling the expressions \eqref{Matrix-Elements-Rx}, \eqref{Matrix-Elements-Ry} for the one-parameter rotations, one finds
\small
\begin{align}
\nonumber
&\sum_{p+q\leqslant N}\BBraket{N}{i,k}{U_{(yz)}(\theta)}{p,q}{N}\BBraket{N}{p,q}{U_{(xz)}(\chi)}{m,n}{N}=\sqrt{\frac{(-1)^{m+n}(N-i)!(N-n)!}{k!n!i!m!(N-i-k)!(N-n-i)!}}
\\
\label{Second}
&\times\frac{(-1)^{i+m}\tan^{k+n}\theta\tan^{i+m}\chi}{\cos^{i-N}\theta\cos^{n-N}\chi\sqrt{(i-N)_{n}(n-N)_{m}}}\,k_{m}(i;\sin^2\chi;N-n)\;k_{n}(k;\sin^2\theta;N-i).
\end{align}
\normalsize
Comparing the expressions \eqref{First} and \eqref{Second}, a short calculation shows that the parameters conspire to give
\begin{align}
\nonumber
&K_{2}(m,n;i,k;\sin^2\chi;\sin^2\theta\cos^2\chi;N)
\\
\label{Trat}
&=\frac{1}{(-N)_{n+m}}\,k_{m}(i;\sin^2\chi;N-n)k_{n}(k;\sin^2\theta;N-i),
\end{align}
as expected from \eqref{Def-Tratnik} and \eqref{Para-Tratnik}. Thus the addition formula \eqref{Addition-Matrix-Elements} leads to the explicit expression for the polynomials $Q_{m,n}(i,k)$ when $R_{12}=0$. 
\subsection{Expansion of the general Krawtchouk polynomials in the Krawtchouk-Tratnik polynomials}
It is possible to find from \eqref{Addition-Matrix-Elements} an expansion formula of the general Krawtchouk polynomials $Q_{m,n}(i,k;N)$ in terms of the Krawtchouk-Tratnik polynomials. To obtain the expansion, one considers the most general rotation $R$, which can be taken of the form
\begin{align*}
R(\phi,\theta,\chi)=R_{(xz)}(\phi)R_{(yz)}(\theta)R_{(xz)}(\chi)=R_{(xz)}(\phi)C(\theta,\chi),
\end{align*}
where $C(\theta,\chi)$ is given by \eqref{Rotation-Tratnik} and where $R_{(xz)}(\phi)$ is as in \eqref{Rotations}. The formula \eqref{Addition-Matrix-Elements} then yields
\begin{align*}
\BBraket{N}{i,k}{U(R(\phi,\theta,\chi))}{m,n}{N}=\sum_{p+q\leqslant N}\BBraket{N}{i,k}{U_{(xz)}(\phi)}{p,q}{N}\;\BBraket{N}{p,q}{U(C(\theta,\chi))}{m,n}{N}.
\end{align*}
The expressions for $\BBraket{N}{p,q}{U((C(\theta,\chi))}{m,n}{N}$ and $\BBraket{N}{i,k}{U_{(xz)}(\phi)}{p,q}{N}$ are given by \eqref{First} and \eqref{Matrix-Elements-Ry}, respectively. Using \eqref{Matrix-Elements} and \eqref{Def-Q} to express the matrix elements of $U(R)$ in terms of the general polynomials $Q_{m,n}(i,k;N)$, one obtains the expansion
\begin{align}
\nonumber
&Q_{m,n}(i,k;N)=\Omega_{i,k;m,n;N}(\phi,\theta,\chi)
\\
\label{Expansion}
&\times \sum_{p=0}^{N-k}\frac{(\tan\phi\sec\theta\tan\chi)^{p}}{p!}\;k_{p}(i;\sin^2\phi;N-k)\,K_2(m,n;p,k;\mathfrak{p}_1,\mathfrak{p}_2;N),
\end{align}
where
$
\mathfrak{p}_1=\sin^2\chi$, $\mathfrak{p}_2=\sin^2\theta\cos^2\chi$,
and
\begin{align*}
&\Omega_{i,k;m,n;N}(\phi,\theta,\chi)\\
&=(-1)^{i}\tan^{i}\phi\cos^{N-k}\phi \left(\frac{C_{33}}{R_{33}}\right)^{N}
\left(\frac{R_{33}}{R_{13}}\right)^{i}
\left(\frac{C_{23}R_{33}}{C_{33}R_{23}}\right)^{k}
\left(\frac{C_{31}R_{33}}{C_{33}R_{31}}\right)^{m}
\left(\frac{C_{32}R_{33}}{C_{33}R_{32}}\right)^{n}.
\end{align*}
Substituting the formula \eqref{Trat} for $K_2(m,n;i,k;\sin^2\chi,\sin^2\theta\cos^2\chi;N)$ and using \eqref{Def-Kr} transforms \eqref{Expansion} in an expression for $Q_{m,n}(i,k;N)$ in terms of hypergeometric series.
\section{Multidimensional case}
We now show in this section how the results obtained thus far can easily be generalized by considering the state vectors of the $d+1$-dimensional harmonic oscillator so as to obtain an algebraic description of the general multivariate Krawtchouk polynomials in $d$ variables that are orthogonal with respect to the multinomial distribution \cite{Iliev-2012}.

Consider the Hamiltonian of the $d+1$-dimensional harmonic oscillator
\begin{equation*}
H = a_1^{\dagger} a_1 + a_2^{\dagger} a_2 + \dots + a_{d+1}^{\dagger} a_{d+1},  \end{equation*}
where the operators $a_i$, $a_i^{\dagger}$ obey the commutation relations \eqref{Weyl} with $i,k=1,\ldots,d+1$. Let $\mathcal{V}_N$ denote the eigensubspaces of the $d+1$-dimensional Hamiltonian $H$ corresponding to the energy eigenvalues $N=0,1,2,\dots$. An orthonormal basis for the space $\mathcal{V}_{N}$ is provided by the vectors
\begin{align}
\label{States}
\kket{n_1,\ldots,n_d}{N}=\ket{n_1,n_2,\ldots,N-n_1-\ldots-n_{d}}.
\end{align}
The action of the operators $a_{i}$, $a_{i}^{\dagger}$ on the basis vectors $\ket{n_1,\ldots,n_{d+1}}$ is identical to the one given in \eqref{Actions}. Since the Hamiltonian of the $(d+1)$-dimensional oscillator is clearly invariant under $SU(d+1)$ and hence $SO(d+1)$ transformations, it follows that the states \eqref{States} provide a reducible representation of the rotation group $SO(d+1)$ in $d+1$ dimensions. 

Let $B$ be a real $(d+1)\times (d+1)$ antisymmetric matrix $(B^{T}=-B)$ and let $R\in SO(d+1)$ be the rotation matrix related to $B$ by $e^{B}=R$. One has evidently $R^{T}R=1$. Consider now the unitary representation
\begin{align}
U(R)=\exp\left(\sum_{j,k=1}^{d+1}B_{jk}a_{j}^{\dagger}a_{k}\right),
\end{align}
which has for parameters the $d(d+1)/2$ independent matrix elements of the matrix $B$. The transformations of the operators $a_{i}^{\dagger}$, $a_{i}$ under the action of $U(R)$ are given by
\begin{align*}
U(R)a_{i}U^{\dagger}(R)=\sum_{k=1}^{d+1}R_{ki}a_{k},\qquad U(R)a_{i}^{\dagger}U^{\dagger}(R)=\sum_{k=1}^{d+1}R_{ki}a_{k}^{\dagger}.
\end{align*}

In the same spirit as in Section 3, one can write the matrix elements of the reducible representations of $SO(d+1)$ on the space $\mathcal{V}_{N}$ of the $(d+1)$-dimensional oscillator eigenstates as follows:
\begin{align}
\label{Matrix-Elements-Multi}
\BBraket{N}{i_1,\ldots,i_{d}}{U(R)}{n_{1},\ldots,n_{d}}{N}=W_{i_1,\ldots,i_{d};N}\,P_{n_1,\ldots,n_{d}}(i_1,\ldots,i_{d};N),
\end{align}
where $P_{0,\ldots,0}(i_1,\ldots,i_{d};N)\equiv1$ and where
\begin{align*}
W_{i_1,\ldots,i_{d};N}=\BBraket{N}{i_{1},\ldots,i_{d}}{U(R)}{0,\ldots,0}{N}.
\end{align*}
It is straightforward to show (as in Section 3) that
\begin{equation*}
\label{W_d}
W_{i_1,\ldots,i_{d};N}={\sqrt{\frac{N!}{i_1! i_2! \dots i_{d}!(N-i_{1}-\ldots-i_{d})!}}} \: R_{1,d+1}^{i_1} R_{2,d+1}^{i_2} \dots R_{d+1,d+1}^{N-i_1-\cdots-i_{d}}.
\end{equation*}
Note that one has
\begin{equation*}
\sum_{i_1+\cdots+i_{d}\leqslant N} W_{i_1,\ldots,i_{d};N}^2 =1, 
\end{equation*} 
which follows immediately from the multinomial formula and from the orthogonality relation
\begin{equation*}
R_{1,d+1}^2+R_{2,d+1}^2+\dots+R_{d+1,d+1}^2=1.
\end{equation*}
It is easily verified that $P_{n_1,\ldots,n_{d}}(i_{1},\ldots,i_{d};N)$ are polynomials in the discrete variables $i_1, \dots, i_d$ that are of total degree $n_1 +\dots + n_d$. These polynomials are orthonormal with respect to the multinomial distribution $W_{i_1,\ldots,i_{d};N}^2$
\begin{equation*}
\sum_{i_1+\cdots+i_{d}\leqslant N}W_{i_1,\ldots,i_{d};N}^2 P_{m_1,\ldots,m_{d}}(i_1,\ldots,i_{d};N) P_{n_1,\ldots,n_{d}} (i_{1},\ldots,i_{d};N)=\delta_{n_1m_1}\cdots \delta_{n_{d}m_{d}}.
\end{equation*}
For the monic polynomials $Q_{n_1,\ldots,n_{d}}(i_1,\ldots,i_{d};N)$, one finds for the generating function
\begin{align*}
&\big(1+\sum_{k=1}^{d}z_k\big)^{N-i_1-\cdots-i_{d}}\prod_{j=1}^{d}\Big(1+\sum_{k=1}^{d}u_{j,k}z_{k}\Big)^{i_j}
\\
&=\sum_{n_1+\cdots+ n_d\leqslant N}\binom{N}{n_1,\ldots,n_{d}}\,Q_{n_1,\ldots,n_{d}}(i_1,\ldots,i_{d};N)\,z_1^{n_1}\cdots z_{d}^{n_{d}},
\end{align*}
where $\binom{N}{x_1,\ldots,x_{d}}$ are the mutinomial coefficients
\begin{align*}
\binom{N}{x_1,\ldots,x_d}=\frac{N!}{x_1!\cdots x_{d}!(N-x_1-\cdots-x_{d})!},
\end{align*}
and where
\begin{align*}
u_{j,k}=\frac{R_{j,k}R_{d+1,d+1}}{R_{j,d+1}R_{d+1,k}}.
\end{align*}
Deriving the properties of these polynomials for a general $d$ can be done exactly as for $d=2$.
\section{Conclusion}
To summarize we have considered the reducible representations of the rotation group $SO(d+1)$ on the energy eigenspaces of the $(d+1)$-dimensional harmonic isotropic oscillator. We have specialized for the most our discussion to $d=2$ with the understanding that it extends easily. We have shown that the multivariate Krawtchouk polynomials arise as matrix elements of these $SO(d+1)$ representations. This interpretation has brought much clarity on the general theory of these polynomials and in particular on the relation to the Krawtchouk-Tratnik polynomials. 

Our main results can equivalently be described as providing the overlap coefficients between two Cartesian bases, one rotated with respect to the other, in which the Schr\"odinger equation for the 3-dimensional harmonic oscillator separates. This paper therefore also adds to studies of interbasis expansions for the harmonic oscillator and more general systems that have been carried out for instance in \cite{Pogo-1998,Pogo-1999,Miller-2011} and references therein.

More results can be expected from this group-theoretic picture, some technical, some of a more insightful nature. In the first category, it is clear that the defining formula \eqref{Matrix-Elements} is bound to yield an expression for the general multivariable Krawtchouk polynomials in terms of single-variable Krawtchouk polynomials and appropriate Clebsch-Gordan coefficients when the rotation group representation spanned by the basis vectors $\kket{m,n}{N}$ in three dimensions for example, is decomposed into its irreducible components. This will be the object of a forthcoming publication \cite{Genest-2013}. To illustrate the possibilities in the second category, let us observe that the analysis presented here puts the properties of the multivariable Krawtchouk polynomials in an interesting light if one has in mind generalizations. One can see for instance a path to a $q$-extension of the multivariate Krawtchouk polynomials. Moreover, understanding that Lie groups will no longer enter the picture in all likelihood, the analysis offers nevertheless an interesting starting point to explore multivariate analogs of the higher level polynomials in the Askey tableau with more parameters than those defined by Tratnik. We hope to report on these related questions in the near future.
\section*{Acknowledgements}
A.Z. thanks the Centre de Recherches Math\'ematiques (CRM) for its hospitality during the course of this work. The research of L.V. is supported in part by the Natural Sciences and Engineering Council of Canada (NSERC). V.X.G. holds an Alexander-Graham-Bell fellowship from NSERC.
\pagebreak
\appendix
\section{Background on multivariate Krawtchouk polynomials}
In order to make the paper self-contained, we shall collect in this appendix a number of properties of the multivariate Krawtchouk polynomials that can be found in the literature. We shall furthermore indicate what is the relation between the parameters that have been used in these references and the rotation matrix elements that arise naturally in the algebraic model presented here. We shall adopt (for the most) the notation of Iliev \cite{Iliev-2012} in the following. 
\subsection*{$d$-variable Krawtchouk polynomials}
In order to define $d$-variable Krawtchouk polynomials, the set of 4-tuples $(\nu,P,\widetilde{P},\mathcal{U})$ is introduced. Here $\nu$ is a non-zero number and $P$, $\widetilde{P}$, $\mathcal{U}$ are square matrices of size $d+1$ with entries satisfying the following conditions:
\begin{enumerate}
\item $P=\mathrm{diag}(\eta_0,\eta_1,\ldots,\eta_d)$ and $\widetilde{P}=\mathrm{diag}(\widetilde{\eta}_0,\widetilde{\eta}_1,\ldots,\widetilde{\eta}_{d}) $ and $\eta_0=\widetilde{\eta}_0=1/\nu$,
\item $\mathcal{U}=(u_{ij})_{0\leqslant i,j\leqslant d}$ is such that $u_{0,j}=u_{j,0}=1$ for all $j=0,\ldots,d$, i.e.
\begin{align*}
\mathcal{U}=
\begin{pmatrix}
1 & 1 & 1 & \cdots & 1\\
1 & u_{1,1} & u_{1,2} & \cdots & u_{1,d}\\
\vdots & & & &\\
1 & u_{1,d} & u_{2,d}& \cdots & u_{d,d}
\end{pmatrix},
\end{align*}
\item The following matrix equation holds
\begin{align}
\label{Annex-Ortho}
\nu P \mathcal{U}\widetilde{P}\mathcal{U}^{T}=I_{d+1}.
\end{align}
\end{enumerate}
It follows from this definition that
\begin{align*}
\sum_{j=0}^{d}\eta_{j}=\sum_{j=0}^{d}\widetilde{\eta}_{j}=1.
\end{align*}
Take $N$ to be a positive  integer and let $m=(m_1,\ldots,m_{d})$ and $\widetilde{m}=(\widetilde{m}_1,\ldots,\widetilde{m}_{d})$ with $m_i$, $\widetilde{m}_i$, $i=1,\ldots,d$, non-negative integers such that $m_1+m_2+\cdots+m_{d}\leqslant N$, $\widetilde{m}_1+\widetilde{m}_2+\cdots+\widetilde{m}_{d}\leqslant N$. Following Griffiths \cite{Griffiths-1971}, the polynomials $Q(m,\widetilde{m})$ in the variables $\widetilde{m}_i$ with degrees $m_i$ are obtained from the generating function
\begin{align*}
\prod_{i=0}^{d}\Big(1+\sum_{j=1}^{d}u_{i,j}z_{j}\Big)^{\widetilde{m}_i}=\sum_{m_1+\cdots m_d\leqslant N}\frac{N!}{m_0!m_1!m_2!\cdots m_d!}Q(m,\widetilde{m})z_1^{m_1}\cdots z_{m}^{m_{d}},
\end{align*}
where $m_0=N-m_1-m_2-\cdots m_d$ and $\widetilde{m}_0=N-\widetilde{m}_1-\widetilde{m_2}-\cdots \widetilde{m}_d$. 

Identifying $(n_1,\ldots,n_{d})$ with $(m_1,\ldots,m_{d})$ and $(i_1,\ldots,i_{d})$ with $(\widetilde{m}_1,\ldots,\widetilde{m}_d)$, the polynomials $P_{n_1,\ldots,n_{d}}(i_{1},\ldots,i_{d};N)$ introduced in \eqref{Matrix-Elements-Multi} are the polynomials $Q(m,\widetilde{m})$ up to a normalization factor.

An explicit formula for $Q(m,\widetilde{m})$ in terms of Gel'fand-Aomoto series has been given by Mizukawa and Tanaka \cite{Tanaka-2004}:
\begin{align}
\label{Annex-Explicit}
Q(m,\widetilde{m})=\sum_{\{a_{ij}\}}\frac{\prod_{j=1}^{d}(-m_{j})_{\sum_{i=1}^{d}a_{ij}}\prod_{i=1}^{d}(-\widetilde{m}_i)_{\sum_{j=1}^{d}a_{i,j}}}{(-N)_{\sum_{i,j}^{d}a_{i,j}}}\,\prod_{i,j=1}^{d}\frac{\omega_{i,j}^{a_{i,j}}}{a_{i,j}!},
\end{align}
where $\omega_{ij}=1-u_{ij}$, $a_{ij}$ are non-negative integers such that $\sum_{i,j=1}^{d}a_{i,j}\leqslant N$.

Let 
\begin{align}
\label{S}
S_1^{2}=\nu P,\qquad S_2^{2}=Q,
\end{align}
and set
\begin{align}
\label{V}
V=S_{1}\mathcal{U}S_{2}.
\end{align}
It then follows that
\begin{align*}
VV^{T}=S_{1}\mathcal{U}S_{2}S_{2}\mathcal{U}^{T}S_1=S_{1}\mathcal{U}Q\mathcal{U}^{T}S_{1}=1,
\end{align*}
$V$ is thus an orthogonal matrix and one has $\det V=\pm 1$. By an appropriate choice of the signs of the entries of the matrices $S_1$ and  $S_2$, one can ensure that $\det(V)=1$ so that $V$ corresponds to a proper rotation. Consequently, the rotation matrix $R$ providing the parameters for the general polynomials $Q_{m,n}(i,k;N)$ in our picture can be obtained from $V$ by rearranging the rows and columns.
\subsection*{Bivariate case}
When $d=2$, the above formulas have been specialized as follows. The matrix elements $u_{ij}$ have been taken \cite{Terwi-2012} to be parametrized by four numbers $p_1$, $p_2$, $p_3$, $p_4$ according to
\begin{subequations}
\label{Annex-uvsp}
\begin{align}
u_{11}&=1-\frac{(p_1+p_2)(p_1+p_3)}{p_1(p_1+p_2+p_3+p_4)}=\frac{p_1p_4-p_2p_3}{p_1(p_1+p_2+p_3+p_4)},\\
u_{12}&=1-\frac{(p_1+p_2)(p_2+p_4)}{p_2(p_1+p_2+p_3+p_4)}=\frac{p_2p_3-p_1p_4}{p_2(p_1+p_2+p_3+p_4)},\\
u_{21}&=1-\frac{(p_1+p_3)(p_3+p_4)}{p_3(p_1+p_2+p_3+p_4)}=\frac{p_2p_3-p_1p_4}{p_3(p_1+p_2+p_3+p_4)},\\
u_{22}&=1-\frac{(p_2+p_4)(p_3+p_4)}{p_4(p_1+p_2+p_3+p_4)}=\frac{p_1p_4-p_2p_3}{p_4(p_1+p_2+p_3+p_4)},
\end{align}
\end{subequations}
with $\eta_i$ and $\widetilde{\eta}_i$ given by
\begin{subequations}
\label{Annex-eta}
\begin{align}
\eta_1&=\frac{p_1p_2(p_1+p_2+p_3+p_4)}{(p_1+p_2)(p_1+p_3)(p_2+p_4)},\quad \eta_2=\frac{p_3p_4(p_1+p_2+p_3+p_4)}{(p_1+p_3)(p_2+p_4)(p_3+p_4)},\\
\widetilde{\eta}_1&=\frac{p_1p_3(p_1+p_2+p_3+p_4)}{(p_1+p_2)(p_1+p_3)(p_3+p_4)},\quad
\widetilde{\eta}_{2}=\frac{p_2p_4(p_1+p_2+p_3+p_4)}{(p_1+p_2)(p_2+p_4)(p_3+p_4)},
\end{align}
\end{subequations}
and where $\eta_0=\widetilde{\eta}_0=1-\eta_1-\eta_2$. The numbers $p_1,\cdots, p_4$ are assumed to be arbitrary apart for certain combinations that would lead to divisions by $0$. It is checked that \eqref{Annex-Ortho} is satisfied with these definitions. It is also observed that these parameters are defined up to an arbitrary common factor since the quadruplet $(\gamma p_1,\gamma p_2,\gamma p_3,\gamma p_4)$ with an arbitrary non-zero $\gamma$ ($\gamma\neq 0$) leads to the same $u_{ij}$ as $(p_1,p_2,p_3,p_4)$. This means that only three of the four parameters $p_i$ are independent. This is related to the fact that 3-dimensional rotations depend at most on 3 independent parameters like the Euler angles for instance. The explicit formula \eqref{Annex-Explicit} thus reduces to
\begin{align}
\nonumber
Q_{m,n}(\widetilde{m},\widetilde{n})=\sum_{i+j+k+\ell \leqslant N}&\frac{(-m)_{i+j}(-n)_{k+\ell}(-\widetilde{m})_{i+k}(-\widetilde{n})_{j+\ell}}{i!j!k!\ell!(-N)_{i+j+k+\ell}}\\
\label{Annex-Explicit-II}
&\times (1-u_{11})^{i}(1-u_{21})^{j}(1-u_{12})^{k}(1-u_{22})^{\ell}.
\end{align}
As for the generating function, it becomes
\begin{align}
\label{Annex-Generating}
(1&+u_{11} z_1+u_{12}z_{2})^{\widetilde{m}_1}(1+u_{21} z_1+u_{22}z_{2})^{\widetilde{m}_2}(1+z_1+z_{2})^{\widetilde{m}_0}\nonumber\\
&=\sum_{m_1+m_2\leqslant N}\frac{N!}{m_0!m_1!m_2!}Q(m,\widetilde{m})z_{1}^{m_1}z_{2}^{m_2}.
\end{align}
From the identification \eqref{Para-1} that brought the generating function \eqref{Gen-Q} into the form \eqref{Annex-Generating}, we can express the parameters $p_1,\ldots,p_4$ in terms of the rotation matrix elements. One observes from \eqref{Annex-uvsp} that
\begin{subequations}
\label{Annex-Identification}
\begin{align}
\frac{u_{11}}{u_{12}}&=-\frac{p_1}{p_2}=\frac{R_{11}R_{32}}{R_{31}R_{12}},\quad 
\frac{u_{21}}{u_{22}}=-\frac{p_4}{p_3}=\frac{R_{21}R_{32}}{R_{31}R_{22}},\\
\frac{u_{12}}{u_{21}}&=\frac{p_3}{p_2}=\frac{R_{12}R_{23}R_{31}}{R_{13}R_{32}R_{21}},\quad
\frac{u_{11}}{u_{22}}=\frac{p_4}{p_1}=\frac{R_{11}R_{23}R_{31}}{R_{13}R_{31}R_{21}},
\end{align}
\end{subequations}
whence it is seen that one possible identification satisfying \eqref{Annex-Identification} is
\begin{align}
\label{Annex-iden}
p_1=\frac{R_{31}}{R_{11}},\quad p_2=-\frac{R_{32}}{R_{12}},\quad p_3=-\frac{R_{23}R_{31}}{R_{13}R_{21}},\quad p_4=\frac{R_{32}R_{23}}{R_{13}R_{22}}.
\end{align}
One can use the affine latitude noted above to write down a more symmetric parametrization (by multiplying the above parameters by $R_{31}$) where
\begin{align*}
p_1=\frac{R_{31}R_{13}}{R_{11}},\quad p_2=-\frac{R_{32}R_{13}}{R_{12}},\quad p_3=-\frac{R_{23}R_{31}}{R_{21}},\quad p_4=\frac{R_{32}R_{23}}{R_{22}}.
\end{align*}

The expressions for $\eta_1$, $\eta_2$, $\widetilde{\eta}_{1}$ and $\widetilde{\eta}_{2}$ in terms of the rotation matrix elements $R_{ij}$ can also be determined. For instance, one obtains from \eqref{Annex-eta} with the help of \eqref{Annex-uvsp}, \eqref{Para-1} and \eqref{Annex-Identification}
\begin{align}
\label{Annex-Vlan}
\eta_1=\frac{p_2}{p_2+p_4}\frac{1}{1+u_{11}}=\frac{R_{13}^2R_{22}R_{31}}{(R_{13}R_{22}-R_{12}R_{23})(R_{13}R_{31}-R_{11}R_{33})}.
\end{align}
Now given that $\mathrm{det}\,R=1$ and $R^{-1}=R^{T}$, one has $R^{T}=\mathrm{adj}\,R$. This yields in particular,
\begin{align*}
R_{31}=R_{12}R_{23}-R_{13}R_{22},\quad 
R_{22}=R_{11}R_{33}-R_{13}R_{31},
\end{align*}
from where we find from \eqref{Annex-Vlan} that
\begin{align*}
\eta_1=R_{13}^2.
\end{align*}
Similarly one arrives at
\begin{align*}
\eta_{2}=R_{23}^2,\quad \widetilde{\eta}_1=R_{31}^2,\quad \widetilde{\eta}_{2}=R_{32}^2,
\end{align*}
and $\eta_0=\widetilde{\eta}_0=1-\eta_{1}-\eta_{2}=R_{33}^2$. This is in keeping with the fact that the polynomials $Q(m,\widetilde{m})$ are orthogonal with respect to the trinomial weight distribution
\begin{align*}
\omega(\widetilde{m}_1,\widetilde{m}_2)=\frac{N!}{\widetilde{m}_1!\widetilde{m_2}!(N-\widetilde{m}_1-\widetilde{m}_2)!}\eta_1^{\widetilde{m}_1}\eta_2^{\widetilde{m}_2}(1-\eta_1-\eta_2)^{N-\widetilde{m}_1-\widetilde{m}_2},
\end{align*}
that should be compared with formula \eqref{Weight}.

In the multivariate case, the parameters $u_{ij}$ of the matrix $\mathcal{U}$ are related to those of $SO(d+1)$ rotations according to formulas \eqref{S} and \eqref{V}. One may verify these relations in the bivariate case with
\begin{align*}
S_1=\mathrm{diag}\,(R_{33},R_{13},R_{23}),\quad
S_2=\mathrm{diag}\,(R_{33},R_{31},R_{32}),\quad
\nu=R_{33},
\end{align*}
and the entries of $\mathcal{U}$ given by \eqref{Para-1}. A direct calculation shows that
\begin{align*}
V=\frac{1}{\nu}S_{1}\mathcal{U}S_{2}=
\begin{pmatrix}
R_{33} & R_{31} & R_{32}\\
R_{13} & R_{11} & R_{12}\\
R_{23} & R_{21} & R_{22}
\end{pmatrix},
\end{align*}
which is the transpose of $R$ after a cyclic permutation of the rows and columns. 
\subsection*{Krawtchouk-Tratnik polynomials}
\normalsize
Recall that the Krawtchouk-Tratnik polynomials $K_{2}(m,n;i,k;\mathfrak{p}_1,\mathfrak{p}_2;N)$ were seen to be a special case of the Krawtchouk polynomials in two variables when $R_{12}=0$ (or $R_{21}=0$ as a matter of fact). We note from \eqref{Annex-iden} that the parametrization in terms of $p_1$, $p_2$, $p_3$, $p_4$ becomes singular in these instances. For completeness, let us record here the recurrence relations that are satisfied by these polynomials. They are obtained directly from the formulas given in Appendix A.3 of \cite{Iliev-2010} and read (suppressing the parameters $\mathfrak{p}_1$, $\mathfrak{p_2}$ and $N$):
\small
\begin{subequations}
\begin{align}
\nonumber
i\,K_2(m,n;i,k)=&\,\mathfrak{p}_1(m+n-N)\Big[ K_2(m+1,n;i,k)-K_2(m,n;i,k)\Big]\\
&+(1-\mathfrak{p}_1)m\Big[K_2(m,n;i,k)-K_2(m-1,n;i,k)\Big],
\label{Tratnik-A}
\\
\nonumber
k\,K_2(m,n;i,k)=&\,\Big[\frac{\mathfrak{p}_1\mathfrak{p}_2}{1-\mathfrak{p}_1}m+\frac{(1-\mathfrak{p}_1-\mathfrak{p}_2)}{1-\mathfrak{p}_1}n+\mathfrak{p}_2(N-m-n)\Big]K_2(m,n;i,k),
\\
\nonumber
&-\frac{\mathfrak{p}_2}{1-\mathfrak{p}_1}m K_2(m-1,n+1;i,k)-\mathfrak{p}_1\frac{(1-\mathfrak{p}_1-\mathfrak{p}_2)}{1-\mathfrak{p}_1}\,n\,K_2(m+1,n-1;i,k)
\\
\label{Annex-Recu-2}
&+\mathfrak{p}_2 \,m\,K_2(m-1,n;i,k)+\frac{\mathfrak{p}_1\mathfrak{p}_2}{1-\mathfrak{p}_1}(N-n-m)K_2(m+1,n;i,k)
\\
\nonumber
& -(1-\mathfrak{p}_1-\mathfrak{p}_2)\,n\,K_2(m,n-1;i,k)-\frac{\mathfrak{p}_2}{1-\mathfrak{p}_1}(N-n-m)K_2(m,n+1;i,k).
\end{align}
\end{subequations}
\normalsize
To check that \eqref{Recurrence-Q-2} reduces to \eqref{Annex-Recu-2} when $R_{12}=0$, given \eqref{Para-Tratnik}, one uses the relations that correspond to $R^{T}R=1$ in this case, namely
\begin{align*}
&R_{11}R_{21}+R_{13}R_{23}=R_{11}R_{31}+R_{13}R_{33}=0,&&R_{11}^2+R_{13}^2=R_{22}^2+R_{32}^2=1 ,\\
&R_{22}R_{21}+R_{32}R_{31}=R_{22}R_{23}+R_{32}R_{33}=0,&&
R_{21}^2+R_{22}^2+R_{23}^2=R_{13}^2+R_{23}^2+R_{33}^2=1,
\end{align*}
to find for instance that
\begin{align*}
&R_{32}^2=\frac{1-\mathfrak{p}_2}{1-\mathfrak{p}_1},\quad R_{22}^2=\frac{1-\mathfrak{p}_1-\mathfrak{p}_2}{1-\mathfrak{p}_2},\quad R_{21}^2=\frac{\mathfrak{p}_1\mathfrak{p}_2}{1-\mathfrak{p}_1},\\
&R_{11}^2=1-\mathfrak{p}_1,\quad R_{33}^2=1-\mathfrak{p}_1-\mathfrak{p}_2,\quad R_{31}^2=\mathfrak{p}_1\frac{(1-\mathfrak{p}_1-\mathfrak{p}_2)}{1-\mathfrak{p}_1},
\end{align*}
and to see that all the factors in \eqref{Recurrence-Q-2} simplify then to those of \eqref{Annex-Recu-2}.

\section*{References}

\end{document}